\documentclass[usenatbib,usegraphicx]{mn2e}
\title[The nature of the Na I D-lines in the Red Rectangle]{The nature of the Na I D-lines in the Red Rectangle}

\author[Joshua D. Thomas et al.]{Joshua~D.~Thomas,$^{1}$\thanks{E-mail:joshua.thomas@utoledo.edu} Adolf~N.~Witt,$^{1}$ Jason~P.~Aufdenberg,$^{2}$ J.~E.~Bjorkman,$^{1}$ \newauthor Julie~A.~Dahlstrom,$^{3}$ S.~R.~Federman,$^{1,6}$ L.~M.~Hobbs,$^{4}$ Uma~P.~Vijh$^{1}$ \newauthor and Donald~G.~York$^{5}$\\
$^{1}$Ritter Astrophysical Research Center, The University of Toledo, Toledo, OH 43606\\
$^{2}$Physical Sciences Department, Embry-Riddle Aeronautical University, Daytona Beach, FL 32114\\
$^{3}$Department of Physics and Astronomy, Carthage College, Kenosha, WI 53140\\
$^{4}$The University of Chicago, Yerkes Observatory, Williams Bay, WI 53191\\
$^{5}$Department of Astronomy \& Astrophysics and The Enrico Fermi Institute, University of Chicago, Chicago, IL 60637\\
$^{6}$Guest Observer, McDonald Observatory, University of Texas at Austin, Austin, TX 78712}

\begin{document}

\date{Accepted for publication in the MNRAS 2011 July 13.  Received 2011 July 12; in original form 2010 December 1 \\The definitive version is available at www.blackwell-synergy.com}

\pagerange{\pageref{firstpage}--\pageref{lastpage}} \pubyear{2011}

\maketitle

\label{firstpage}

\begin{abstract}

In this paper we examine the profiles of the complex Na\,\textsc{i}~D-lines in the Red Rectangle. The spectra were acquired with the ARCES \'{e}chelle spectrograph ($R = 38,000$) on the 3.5-m telescope at the Apache Point Observatory. Additional spectra taken with STIS were acquired from the Hubble Legacy Archive (HLA) and were used to independently confirm the spatial origin of the spectral features. The profile of a single D-line consists of double-peaked emission, red-shifted absorption and blue-shifted absorption. We find that the double-peaked emission originates from the bipolar outflow, the red-shifted absorption feature is due to the photospheric line, and the blue-shifted absorption arises from the bipolar outflow as seen against the photosphere of the luminous post-AGB component in HD~44179. In order to better understand the Na\,\textsc{i}~D-line profile, we examined the periodically variable asymmetric photospheric absorption lines.  The asymmetric lines are interpreted as a signature of slow self-accretion following enhanced mass-loss around periastron.  An empirical model was constructed to remove the photospheric component from the Na\,\textsc{i}~D-line profile in order to study the nebular emission and absorption of sodium along the line-of-sight to the primary. This paper also discusses the different origins of the single-peaked emission, the double-peaked emission and the blue-shifted and red-shifted absorption components.

\end{abstract}


\section{Introduction}

The Red Rectangle (RR), an enigmatic protoplanetary nebula, reveals itself to be a complex astrophysical system.  The central source of the nebula, HD~44179, is a binary system in a short-lived evolutionary phase between the asymptotic giant branch (AGB) and the planetary nebula stage of its life, when the expelled circumstellar matter is ionized by the photometric primary star's hot core \citep{vanwinckel2003, vanwinckel2007, siodmiak2008, szczerba2007}.  The dominant features of the nebula's form are the $\times$-shaped biconical outflows (nearly 2{\arcmin} on the sky \citep{cohen2004}) that emerge from a disc (roughly 0.1{\arcsec} on the sky), seen nearly edge-on.      We will refer to this optically-thick disc, resolved by \textit{HST} \citep{cohen2004} and ground-based interferometric observations \citep{roddier1995, osterbart1997}, as the circumbinary disc.  The dimensions assumed  for the circumbinary disc are based on a distance of 710~pc \citep{menshchikov2002}.  While the distance is uncertain, we adopt this value in our investigation.  \citet{bujarrabal2005} traced the outer radius of the circumbinary disc to 1,850~AU.  The radius of the central cavity in the circumbinary disc is obtained from a scattered light model to be 14~AU \citep{menshchikov2002}.  The thickness of the circumbinary disc has been determined to be between 90~AU \citep{menshchikov2002} and 107~AU \citep{bujarrabal2005, roddier1995, osterbart1997}.  It is worth noting that values of the viewing angle vary from $\sim$ 0$\degr$ to 11$\degr$ \citep{lopez1995,roddier1995,menshchikov2002}.  These viewing angles would correspond to inclination angles between $\sim$ 90$\degr$ to 79$\degr$.  \citet{bujarrabal2005} find that an intermediate value of 5$\degr$ fits their data best, which would correspond to an inclination angle of 85$\degr$.  

At the centre of the cavity in the circumbinary disc are the components of the binary system: the photometric primary (a mass-losing post-AGB star), and the secondary (a main sequence star with an accretion disc and high-velocity jet).  For simplicity we shall refer to the photometric primary as the primary star throughout this paper.  The primary star has an effective temperature of 7,700~K, an estimated mass of 0.8~M$_{\odot}$, a luminosity of about 6,000~L$_{\odot}$ \citep{menshchikov2002}, and a radius of 0.21~AU \citep{witt2009}.  The secondary has a mass of roughly 0.9~M$_{\odot}$ and is believed to be a main sequence star \citep{witt2009}.  In the work of \citet{witt2009} the dependence of the H$\alpha$ line on orbital phase was investigated, revealing a high velocity jet ($v_{max} \sim~560$~km~s$^{-1}$) originating from an accretion disc around the secondary.  The accretion disc around the secondary is powered by mass transfer from the primary star via Roche-lobe overflow at periastron \citep{witt2009}.  The ionizing photons from the hot inner regions of the accretion disc provide the necessary energy to excite the Extended Red Emission (ERE).  The ERE requires far-ultraviolet photons with energies E $>$ 10.5 eV \citep{witt2006}.   The accretion disc also provides the energy source to maintain the H\,\textsc{ii} region inside the cavity of the optically-thick circumbinary disc that was suggested by \citet{jura1997}.

\begin{figure}

\includegraphics[width=80mm]{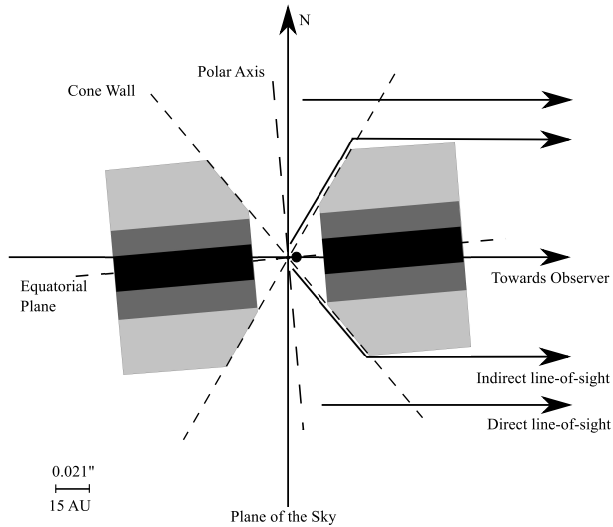}

\caption{This to-scale diagram of the inner region of the RR was constructed for the distance of 710~pc.  The outer radius of the circumbinary disc, 1,850~AU, is not shown.  The central binary is displayed a factor of 10 larger than the scale for the rest of the system.  The inclination angle, angle between the plane of the sky and the equatorial plane, is 85$\degr$.  The direct and indirect lines-of-sight are also indicated for both the northern and southern ends of the nebula.  The cavity is filled with the low-velocity, mass-loss driven outflow of the primary star and the high-velocity jet from the accretion disc of the secondary.  The location of the primary star is shown for the orbital phase of inferior conjunction; see Fig.~\ref{fig2} for a to-scale diagram of the binary orbits depicting the orbital phases. \label{fig1}}
\end{figure}

The near edge-on optically-thick circumbinary disc prevents a direct line-of-sight view of the binary.  Light originating inside the cavity can only be observed via scattering off material located above and below the disc.  This obscured view of the binary, illustrated in Fig.~\ref{fig1}, will be referred to as the indirect line-of-sight and is responsible for the difficult interpretation of the spectra.  The indirect line-of-sight to the central object was first studied by \citet{waelkens1996}, who determined the effective inclination angle to be 35$\degr$.  The effective inclination angle is the angle at which light from the central binary is scattered into our line-of-sight by the material near the top and bottom of the disc.  This viewing geometry has been confirmed by high-resolution optical \citep{osterbart1997, cohen2004} and infrared imaging \citep{tuthill2002}.  For light from the central source (which can only been seen via scattering), we adopt the effective inclination angle of 35$\degr$ in our calculations. 

The outer regions of the nebula are, however, seen directly.  To properly study emission processes that originate in the outer nebula one must use the inclination angle of the nebula (85$\degr$) rather than the effective inclination angle (35$\degr$).

Since HD~44179 can only be studied via scattered light, it is important to discuss the known uncertainties. The effective angle at which we are viewing the stars is likely the mean scattering angle. The variation of the scattering angle over the duration of the orbit, which best fits the 0.14~mag variation in the light curve, is 1.6$\degr$ \citep{waelkens1996}. The average value for the scattering angle corresponds to the effective inclination angle of 35$\degr$. In addition, the tilt of the system implies that the scattering angle for light from the north and the south of the nebula is slightly different (of order 5$\degr$). Therefore, the effective inclination angle is just that, an effective value that we use in order to make progress in understanding this complex system. The variation in the value for the effective inclination angle will cause uncertainties in; the orbital elements, the derived velocities, and the mass of the secondary.

The properties of the stars also depend upon the estimation of the luminosity \citep{menshchikov2002}. The luminosity depends on the uncertain distance to the star, assumptions about the geometry of the circumbinary-disc, and the evolutionary state of the system. Since this a close binary, it is likely that there has been some interaction in the past resulting in a complex evolutionary history. Therefore, it is likely that the evolutionary track of the stars differ from that of single stars, as in \citet{Vassiliadis1994} and \citet{bloecker1995}. As a result, the quoted values that we have used for the mass and luminosity of the primary remain uncertain. At present, these uncertainties can not be addressed with current observational data.

Studying the nebular outflow and its mechanisms is important for understanding how stellar material is returned to the interstellar medium.   The detectable emission of the H$\alpha$ recombination line in our spectra is limited to the inner parts of the nebula.  The H$\alpha$ emission is not observed in the bi-conical lobes far away from the central source.  Studying the Na\,\textsc{i}~D-lines, easily excited resonance lines, allows us to probe the mass-loss to much greater distances from the central source.  The Na\,\textsc{i}~D-lines also allow us to directly study the mass-loss from the primary star.   

The details of the observations, and the radial velocity (RV) measurements are given in {\S}~\ref{sec-observation}.  The determination of the orbital parameters is discussed in {\S}~\ref{sec-orbital}.  In {\S}~\ref{sec-abundance} we discuss the Na\,\textsc{i} abundance and the model atmosphere used to determine the photospheric line profiles.  The behaviour of the photospheric lines and the processing of light in the RR is discussed in {\S}~\ref{sec-lineshape}.  The origins and measurements of the components of the Na\,\textsc{i}~D-line profile are discussed in {\S}~\ref{sec-naprofiles}.  In {\S}~\ref{sec-singlevsdouble} the origin of the single and double peaks are discussed. The conclusions are presented in {\S}~\ref{sec-conclusions}.

\section{Observations and measurement} \label{sec-observation}

\subsection{Ground-based observations} 

\subsubsection{Apache Point Observatory}

\begin{figure}
\includegraphics[width=90mm]{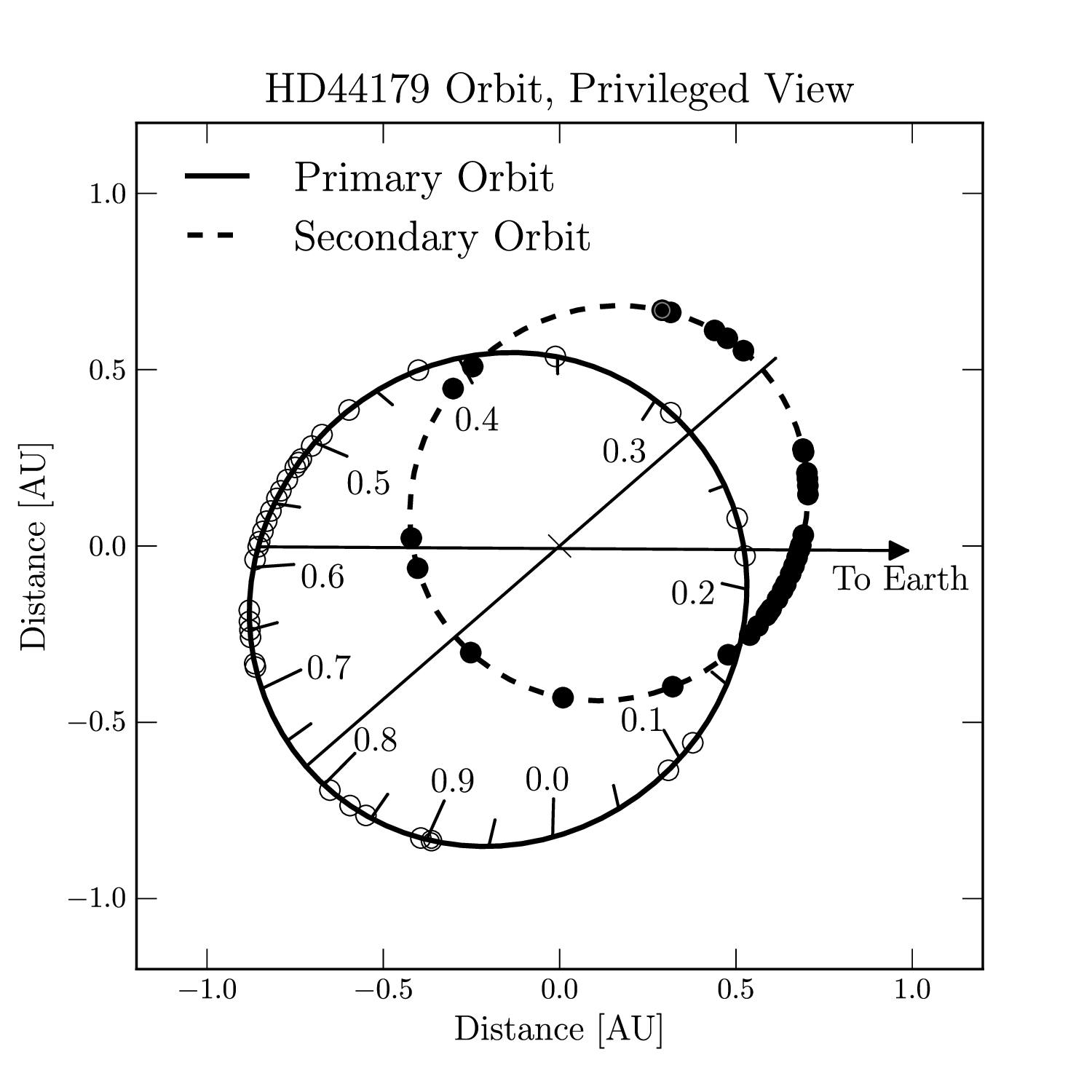}
\caption{Based on the orbital parameters shown in the column M2 of Table~\ref{table2}, the orbits of the binary components have been plotted to-scale (assuming a distance of 710~pc) and are shown from the privileged vantage point above the orbit (inclination i = 0).  The solid ellipse indicates the orbit of the primary, while the dashed ellipse indicates the orbit of the secondary.  The hash at the origin marks the centre-of-mass.  Orbital phases are designated on the primary orbit with the major and minor hashes.  The direction of motion for the primary star is counter clockwise.  Apastron (A) $\phi = 0.79$, and periastron (P) $\phi = 0.29$ occur at the intersection of the orbit and the line passing through the antipodes of both orbits.  The direction to Earth is to the right and 5$\degr$ below the plane of the page.  Inferior conjunction (IC) $\phi = 0.21$, and superior conjunction (SC) $\phi = 0.59$ occur at the intersection of the arrow indicating the direction to Earth and the orbit.  The terminology of IC and SC was chosen to be consistent with \citet{waelkens1996} and \citet{witt2009}.  Maximum red-shift of the primary occurs at $\phi = 0.42$.  The location of phase zero was chosen to agree with \citet{waelkens1996} and corresponds to JD~2448300.  The circles on the orbits indicate the locations of the stars for all 33 APO observations.  All distances are in astronomical units (AU).  Note that the diameter of the post-AGB primary is estimated to be 0.42 AU. \label{fig2}}
\end{figure}

\begin{figure}
\begin{center}
\includegraphics[width=60mm]{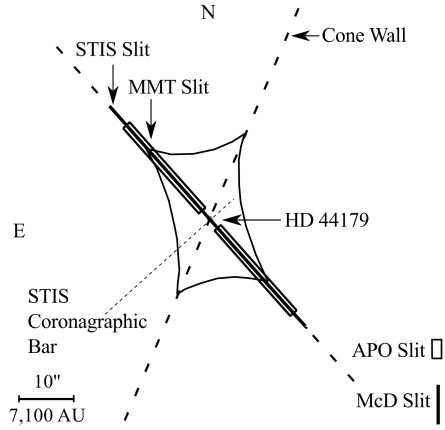}
\end{center}
\caption{A to-scale diagram showing the RR as seen on the sky. The STIS slit is to-scale length wise, but not width wise, being only 0.2{\arcsec} wide.  The width of the coronagraphic bar is to-scale.  The two MMT slit locations of \citet{schmidt1991}, that are mentioned in this paper, are shown to-scale.  The size of the APO slit is shown on the lower right.  The McDonald (McD) slit, also shown on the lower right, is to-scale length wise, but not width wise owing to its small width on the sky (0.34{\arcsec}).  Both the APO and McD slits were centred on HD~44179 during observations, the orientation of these slits on the sky cannot be controlled. \label{fig3}}
\end{figure}

In the present paper we examine 33 high-resolution \'{e}chelle spectra of HD~44179. The data set includes the 17 spectra used in the work of \citet{witt2009}.  The observation dates are detailed in Table~\ref{table1}.  The orbital coverage of the observations is illustrated in Fig.~\ref{fig2}.  The details of Fig.~\ref{fig2} will be discussed in {\S}~\ref{sec-orbital}.  The spectra were obtained with the ARCES \'{e}chelle spectrograph \citep{wang2003} on the 3.5-m telescope at the Apache Point Observatory.  The spectral range of the spectrograph is 3,700~{\AA}  to 10,000~{\AA}  at a resolving power of $R = 38,000$.  The velocity resolution is 8~km~s$^{-1}$.  The entrance slit used measures 1.6{\arcsec}$\times$3.2{\arcsec} on the sky and cannot be rotated.  The relative size of the slit compared to the nebula can be seen in Fig.~\ref{fig3}.  The first spectrum in the combined data set was acquired on 2001 February 6.  The most recent spectrum was acquired on 2009 April 7.  The first 7 and 17 spectra, respectively, were studied previously by \citet{hobbs2004} and \citet{witt2009}.  Almost all the new spectra were taken by DGY, based on prescribed dates from ANW, referring to the results and implications of \citet{witt2009}. The data in Table~\ref{table1} are based on reduction of the raw spectra to normalized spectra by JAD and measurements for individual nights by LMH.  The reduction of the spectra and instrumental set-up are unchanged since the previous work by \citet{thorburn2003} and by \citet{hobbs2004}.  

\begin{figure*}
\includegraphics[width=160mm]{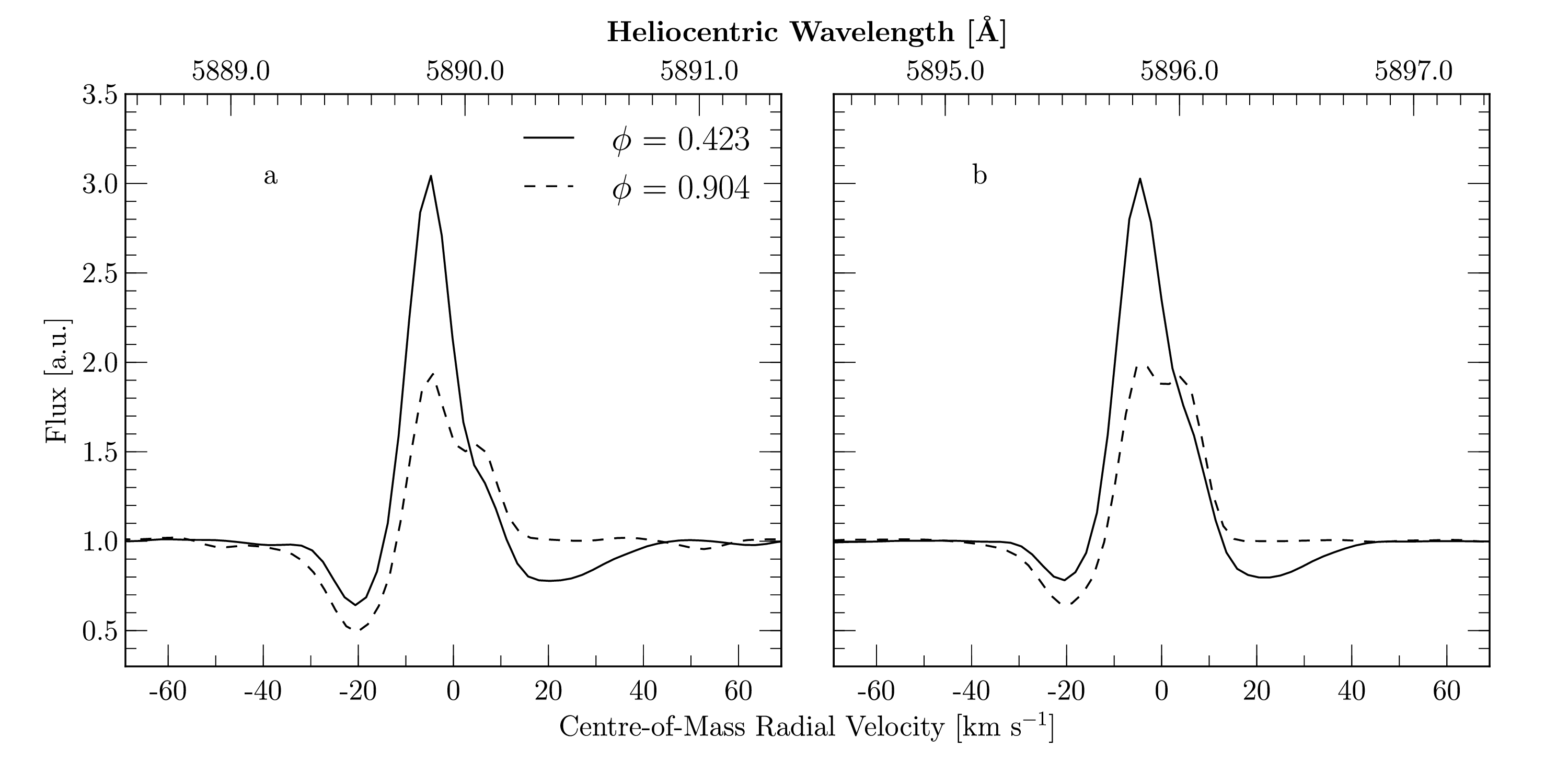}
\caption{The spectra shown were taken at APO with a resolving power of $R = 38,000$.  Panel a shows the Na\,\textsc{i} D2 5889.950~{\AA} line, while panel b shows the D1 5895.924~{\AA} line. The spectra are the observed, averaged, and unsmoothed data with S/N of roughly 300. The spectra are shown at the phases of $\phi = 0.423$ (solid line) and $\phi = 0.904$ (dashed line) corresponding to the maximum and minimum red-shift, respectively, of the absorption associated with the photosphere of the primary star.  See Fig.~\ref{fig2} for the respective locations of the stars in the binary system during these phases.  The flux is in arbitrary units (a.u.).  The heliocentric velocity can be determined by adding 19~km~s$^{-1}$ to the centre-of-mass velocity. \label{fig4}}
\end{figure*}

\begin{figure}
\includegraphics[width=90mm]{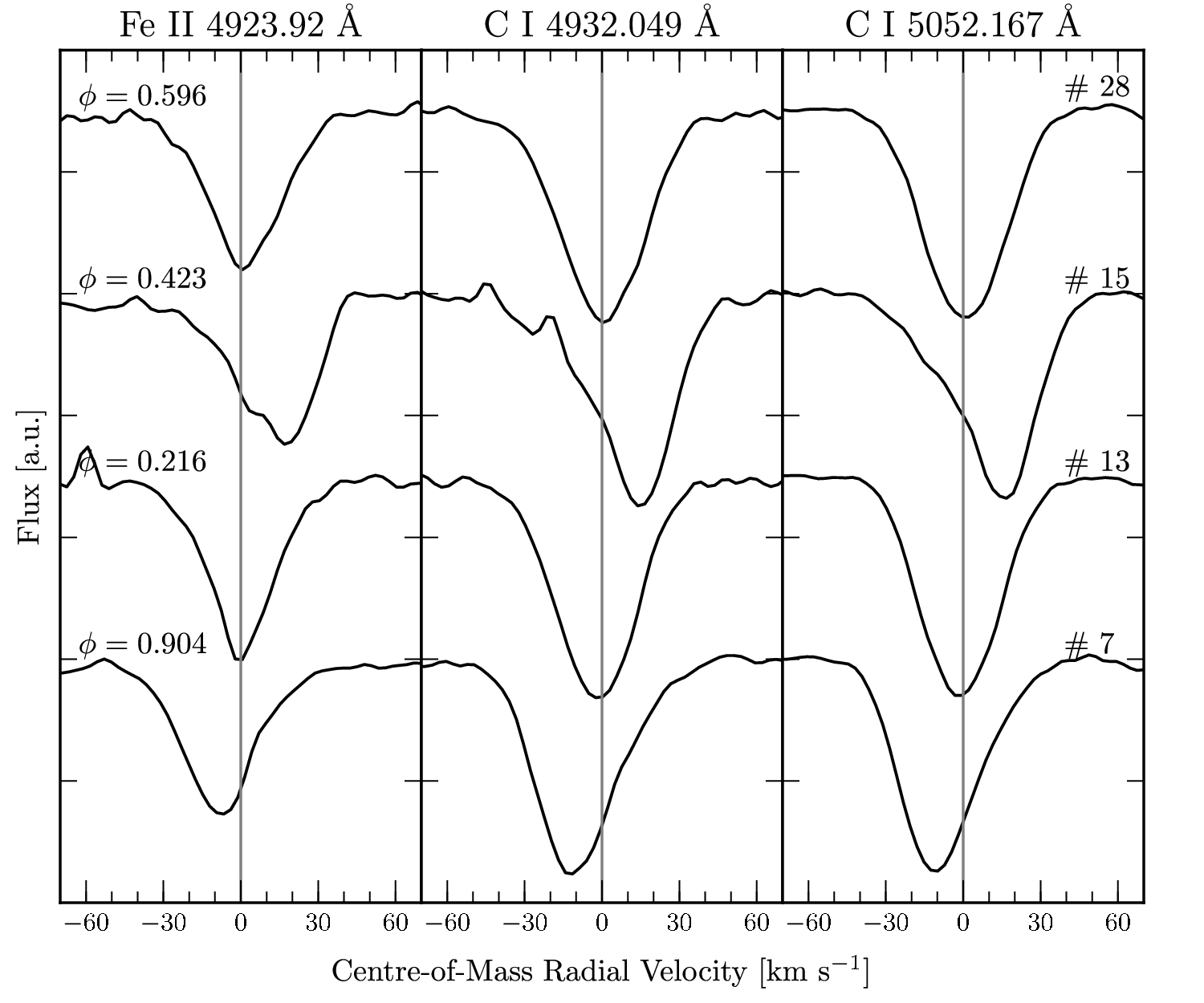}
\caption{Shown is a comparison of three observed, averaged, unsmoothed, photospheric lines.  These data were acquired with APO and have a resolving power of $R = 38,000$.  The C\,\textsc{i} 5052.167~{\AA} profiles were divided by 1.5 in order to show them on the same vertical scale.  Note the similarity between different species, ionization states and wavelengths.  On the far right, each profile is labelled with observation number, while on the far left the phase is shown.  Spectra \#\,13 and \#\,28 occur near inferior and superior conjunction respectively.  Spectrum \#\,7 occurs near minimum red-shift of the primary and \#\,15 occurs at maximum red-shift.  Fig.~\ref{fig2} illustrates the locations of these key orbital positions.  To convert back to heliocentric velocity, add 19~km~s$^{-1}$ (the centre-of-mass velocity of the system). \label{fig5}}
\end{figure}

The present study focuses on the complex Na\,\textsc{i}~D-lines, as exemplified in Fig.~\ref{fig4}.  The large number of observations were needed in order to study variability in the spectra over the course of the binary orbit, not for co-adding.  Each observation is comprised of 2 averaged spectra, with signal-to-noise (S/N) at the continuum in the region of the Na\,\textsc{i}~D-lines of order 300 for the combined spectrum.  We have also examined several absorption lines with no emission component, Fig.~\ref{fig5}.  In our ground-based spectra, the slit (1.6{\arcsec}$\times$3.2{\arcsec}) receives both the scattered light from the primary star and the emission from the inner nebula simultaneously.

\subsubsection{McDonald Observatory}

\begin{figure*}
\includegraphics[width=160mm]{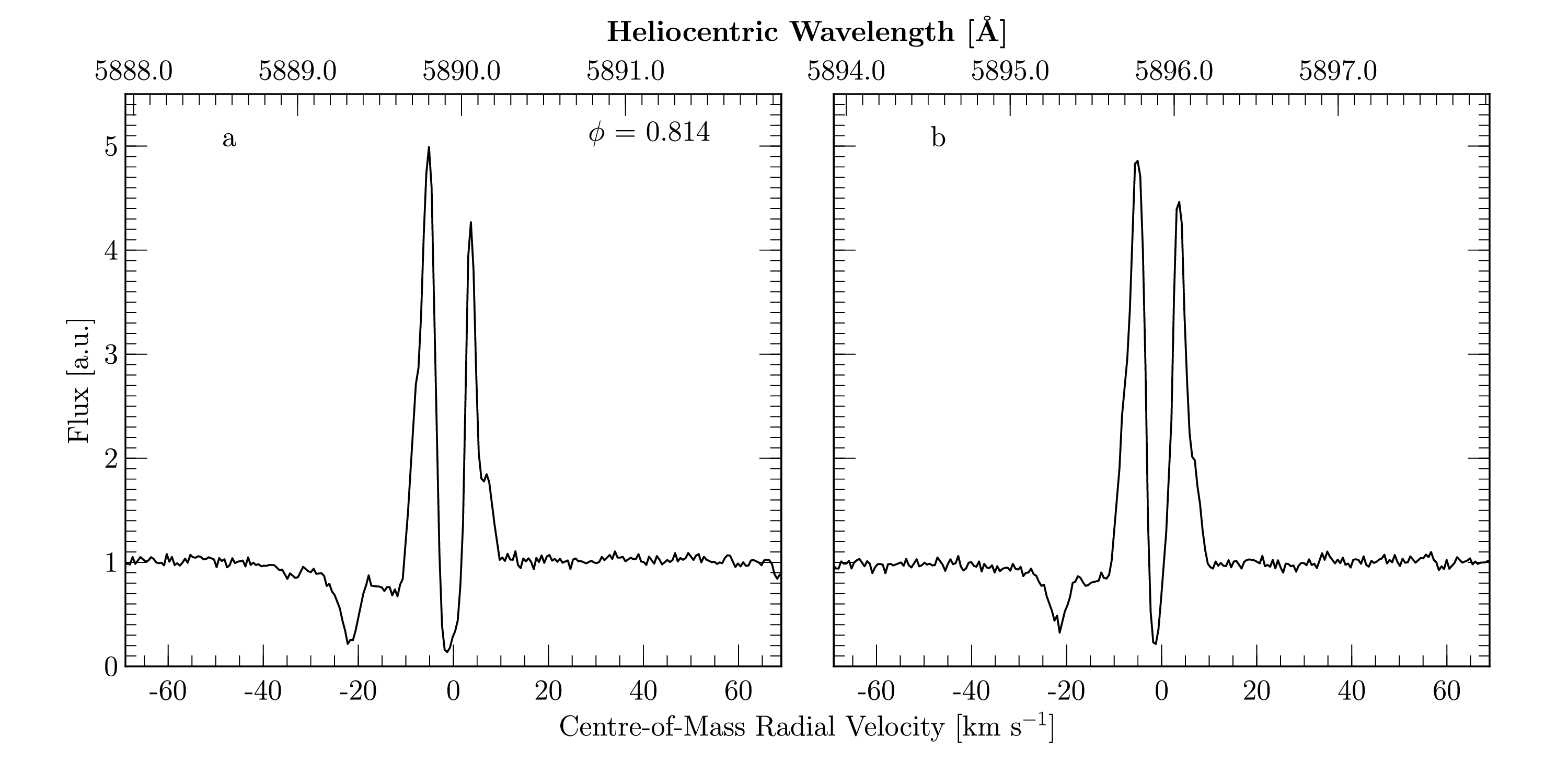}
\caption{The profiles shown were acquired at McDonald Observatory with a resolving power R $\approx$ 200,000.  Panel a shows the Na\,\textsc{i} D2 5889.950~{\AA} line, while panel b shows the D1 5895.924~{\AA} line. The spectra are the observed, averaged, unsmoothed data and have a S/N of approximately 50. The spectra are shown at the phase of $\phi = 0.814$ (Fig.~\ref{fig2}).  The flux is in arbitrary units (a.u.) . The heliocentric velocity can be determined by adding 19~km~s$^{-1}$ to the centre-of-mass velocity. \label{fig6}}
\end{figure*}

A spectrum of the Na\,\textsc{i}~D-lines was also acquired with the Harlan J. Smith 2.7-m telescope at the McDonald Observatory (McD) using the Tull spectrograph \citep{tull1995} in its high-resolution mode (ts21).  The observation was made on 2010 December 24.  The crossed-dispersed echelle spectrometer was configured with the 79~g~mm$^{-1}$ grating (E1), the 145~$\mu$m slit (Slit 2), and a 2048~$\times$~2048 CCD (TK3). The slit is approximately 0.34{\arcsec}$\times$8.2{\arcsec} on the sky.  The slit orientation on the sky is not fixed. Order 39 of the E1 grating spectrum was centred at 5890~{\AA}.  The resolving power (R $\approx$ 200,000) was estimated by measuring the FWHM of the unblended lines in the Th-Ar comparison spectrum.  The corresponding velocity resolution is 1.5~km~s$^{-1}$.  The orbital phase of the observation was $\phi = 0.814$.  The data were reduced and wavelength-calibrated with standard \textsc{iraf} routines.  Plotted in Fig.~\ref{fig6} is the averaged profile for the Na\,\textsc{i} 5889.950~{\AA} line, with a continuum S/N ratio of approximately 50 near the sodium lines.  The average profile was constructed from four 1800~s exposures.

\begin{table*}
\centering
\begin{minipage}{160mm}
\caption{Details of the APO observations and measurements \label{table1}}
\begin{tabular}{lllllllllllll}
\hline
{\#}	&	Year	&	Month	& Day	&	UT	&	HJD	&	$\Delta$HJD	&	Phase\footnote{Based on a period of 318 days; the tabulated phase values are orbit number before the decimal point, followed by the phase value.  The orbit numbers start with this observing campaign.  Phase 0 corresponds to JD 2448300.  Refer to Fig.~\ref{fig2}.}	&	$\Delta$V$_{helio}$\footnote{All velocities are in km~s$^{-1}$.}	&	$\Delta$V$_{LSR}$	&	RV\footnote{RV determined as in \citet{hobbs2004}.}	&	RV$_{strong}$\footnote{RV measured for the strong component of the C\,\textsc{i} 5052~{\AA} line.}	&	RV$_{weak}$\footnote{RV measured for the weak component of the C\,\textsc{i} 5052~{\AA} line.}	\\
\hline
1	&	2001	&	2	&	6	&	4:21:46	&	2451946.68531	&	0	&	1.470	&	-16.4	&	-35.3	&	27.2	&	30.6	&	13.7	\\
2	&	2002	&	12	&	31	&	7:15:39	&	2452639.80723	&	693.12	&	3.650	&	-1.6	&	-20.4	&	17.3	&	14.4	&	23.2	\\
3	&	2003	&	11	&	24	&	8:25:03	&	2452967.85463	&	1021.17	&	4.681	&	13.9	&	-4.9	&	17.3	&	14.4	&	23.1	\\
4	&	2004	&	1	&	3	&	8:15:46	&	2453007.84896	&	1061.16	&	4.807	&	-3.0	&	-21.8	&	11.9	&	9.4	&	16.7	\\
5	&	2004	&	1	&	10	&	4:28:03	&	2453014.69072	&	1068.01	&	4.829	&	-5.6	&	-24.4	&	12.3	&	8.3	&	20.9	\\
6	&	2004	&	1	&	31	&	4:58:31	&	2453035.71118	&	1089.03	&	4.895	&	-14.2	&	-33.0	&	10.7	&	8.1	&	16.1	\\
7	&	2004	&	2	&	3	&	4:21:35	&	2453038.68538	&	1092	&	4.904	&	-15.2	&	-34.0	&	10.7	&	6.5	&	23.3	\\
8	&	2006	&	11	&	12	&	10:33:59	&	2454051.94356	&	2105.26	&	8.090	&	17.7	&	-1.2	&	13.9	&	12.4	&	21.1	\\
9	&	2006	&	12	&	28	&	5:36:38	&	2454097.73848	&	2151.05	&	8.234	&	-0.1	&	-18.9	&	20.4	&	23.0	&	9.5	\\
10	&	2007	&	1	&	15	&	5:49:11	&	2454115.74694	&	2169.06	&	8.291	&	-8.0	&	-26.8	&	25.3	&	28.3	&	10.3	\\
11	&	2007	&	3	&	28	&	1:58:16	&	2454187.58204	&	2240.9	&	8.517	&	-24.8	&	-43.7	&	26.0	&	31.3	&	12.0	\\
12	&	2007	&	10	&	3	&	11:45:44	&	2454376.99043	&	2430.31	&	9.112	&	24.7	&	5.8	&	14.2	&	13.3	&	20.6	\\
13	&	2007	&	11	&	5	&	8:42:34	&	2454409.86573	&	2463.18	&	9.216	&	19.9	&	1.1	&	18.1	&	17.8	&	17.8	\\
14	&	2007	&	12	&	18	&	8:22:37	&	2454452.85368	&	2506.17	&	9.351	&	4.1	&	-14.7	&	29.6	&	33.9	&	11.1	\\
15	&	2008	&	1	&	10	&	5:30:43	&	2454475.73424	&	2529.05	&	9.423	&	-5.7	&	-24.6	&	31.0	&	36.4	&	12.9	\\
16	&	2008	&	2	&	8	&	6:51:57	&	2454504.78953	&	2558.1	&	9.514	&	-17.1	&	-36.0	&	26.2	&	31.4	&	13.1	\\
17	&	2008	&	2	&	29	&	2:18:08	&	2454525.59801	&	2578.91	&	9.580	&	-22.2	&	-41.0	&	21.3	&	18.7	&	27.1	\\
18	&	2008	&	12	&	16	&	7:15:49	&	2454816.80727	&	2870.12	&	10.496	&	4.8	&	-14.0	&	26.0	&	31.8	&	10.5	\\
19	&	2008	&	12	&	19	&	7:34:18	&	2454819.82015	&	2873.13	&	10.505	&	3.4	&	-15.4	&	24.6	&	31.8	&	10.5	\\
20	&	2008	&	12	&	25	&	5:59:41	&	2454825.75448	&	2879.07	&	10.524	&	1.0	&	-17.9	&	24.6	&	30.7	&	8.0	\\
21	&	2008	&	12	&	28	&	5:48:06	&	2454828.74644	&	2882.06	&	10.533	&	-0.4	&	-19.2	&	24.8	&	29.5	&	13.1	\\
22	&	2008	&	12	&	31	&	6:39:41	&	2454831.78225	&	2885.1	&	10.543	&	-1.8	&	-20.6	&	23.3	&	28.6	&	12.0	\\
23	&	2009	&	1	&	2	&	6:28:38	&	2454833.77109	&	2887.09	&	10.549	&	-2.7	&	-21.5	&	22.8	&	11.2	&	18.6	\\
24	&	2009	&	1	&	5	&	7:04:07	&	2454836.79917	&	2890.11	&	10.558	&	-4.1	&	-22.9	&	23.1	&	20.7	&	28.9	\\
25	&	2009	&	1	&	7	&	6:26:23	&	2454838.77271	&	2892.09	&	10.565	&	-4.9	&	-23.7	&	23.2	&	20.7	&	29.3	\\
26	&	2009	&	1	&	10	&	6:17:57	&	2454841.76703	&	2895.08	&	10.574	&	-6.2	&	-25.0	&	22.9	&	21.9	&	28.9	\\
27	&	2009	&	1	&	14	&	6:47:14	&	2454845.78727	&	2899.1	&	10.587	&	-7.9	&	-26.8	&	20.8	&	19.7	&	26.4	\\
28	&	2009	&	1	&	17	&	6:33:11	&	2454848.77743	&	2902.09	&	10.596	&	-9.2	&	-28.0	&	19.8	&	18.2	&	24.4	\\
29	&	2009	&	1	&	30	&	2:14:41	&	2454861.59742	&	2914.91	&	10.636	&	-13.8	&	-32.6	&	18.3	&	16.8	&	30.5	\\
30	&	2009	&	2	&	2	&	1:51:48	&	2454864.58138	&	2917.9	&	10.646	&	-14.8	&	-33.7	&	18.2	&	15.6	&	23.4	\\
31	&	2009	&	2	&	6	&	1:25:21	&	2454868.56281	&	2921.88	&	10.658	&	-16.2	&	-35.0	&	17.0	&	15.6	&	22.8	\\
32	&	2009	&	2	&	13	&	2:40:55	&	2454875.61487	&	2928.93	&	10.680	&	-18.5	&	-37.3	&	17.6	&	15.0	&	21.9	\\
33	&	2009	&	4	&	7	&	2:00:20	&	2454928.58261	&	2981.9	&	10.847	&	-24.4	&	-43.2	&	12.7	&	7.3	&	25.3	\\
\hline
\end{tabular}
\end{minipage}
\end{table*}

\subsection{Space-based observations} 

\subsubsection{HST/STIS}

\begin{figure}
\includegraphics[width=90mm]{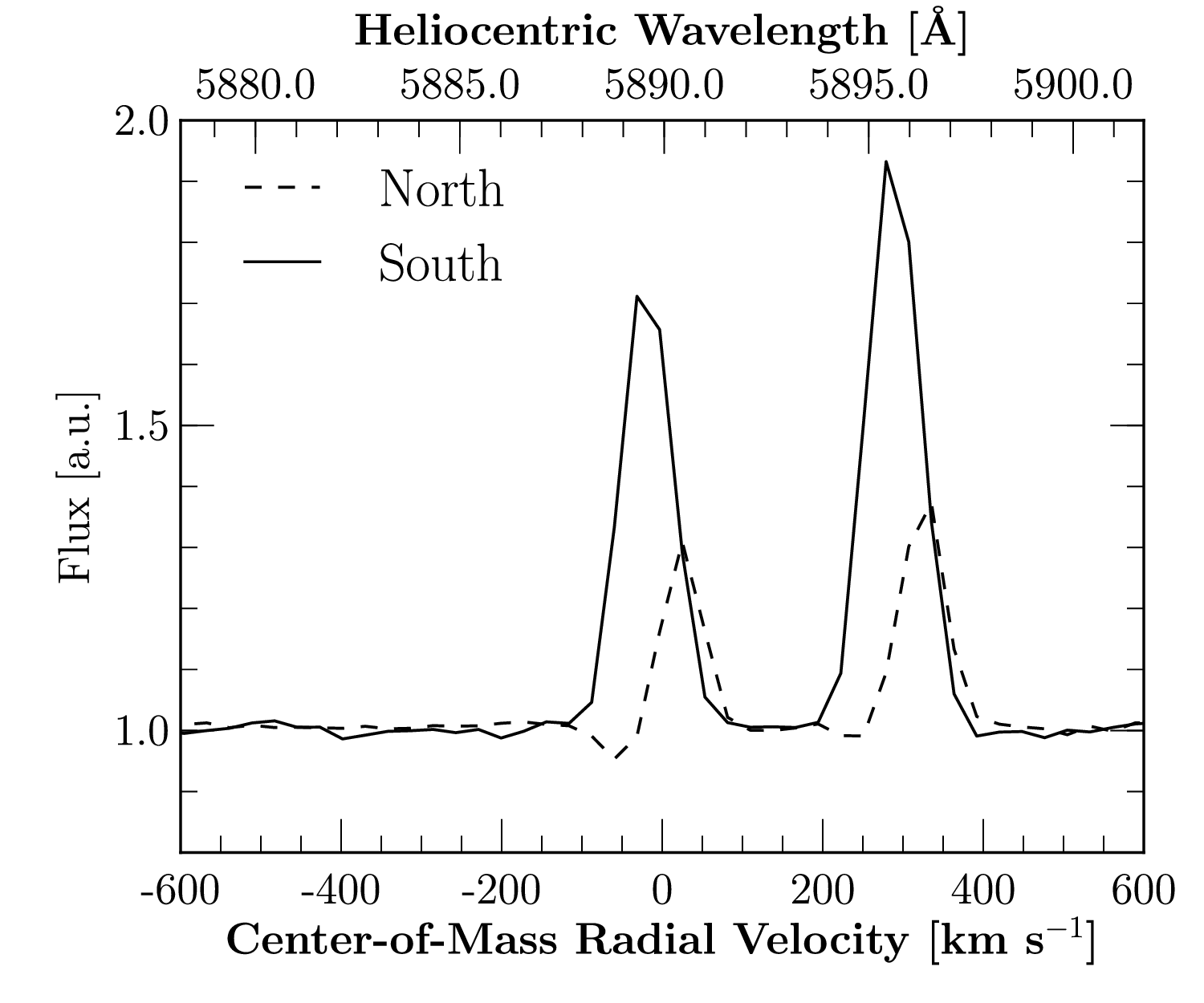}
\caption{This spectrum of the Na\,\textsc{i} D-lines was taken with STIS and has a resolving power $R = 5,000$.  The north and south apertures were extracted separately.  The centre-of-mass velocity scale is centred on the Na\,\textsc{i} D2 5889.950~{\AA} line.  To convert back to heliocentric velocity, add 19~km~s$^{-1}$.  These spectra were acquired during the phase $\phi = 0.19$, see Fig.~\ref{fig2}.  \label{fig7}}
\end{figure}

Na\,\textsc{i}~D spectra obtained with STIS can be seen in Fig.~\ref{fig7}.  These spectra were obtained through the Hubble Legacy Archive.  The spectral resolution of the STIS spectra is lower than that of the other spectra presented in this paper, but they have the superb spatial resolution of the \textit{HST} along the slit.  The 1440~s exposure of the Na\,\textsc{i}~D-lines was taken on 1999 February 12 by Hans van Winckel (proposal ID~7297).  The 52{\arcsec}$\times$0.2{\arcsec} F1 aperture and the G750M grating with a resolving power of $R = 5,000$ were used to acquire this spectrum.  The velocity resolution of these data is 60~km~s$^{-1}$.  The S/N at the continuum for both the sodium and hydrogen spectrum presented in the present paper is of order 200.   The occulting bar measures $0.5${\arcsec} $\times$ 0.2{\arcsec}.  The length of the slit is 52{\arcsec}.  This coronagraphic slit spectrum was oriented along the north-east whisker, position angle 42.056$\degr$.  This slit orientation is similar to one of the slit orientations in \citet{schmidt1991}.  Fig.~\ref{fig3} is a to-scale diagram illustrating the orientation and location of the various slits as projected on to the nebula.  The relevant slits from \citet{schmidt1991}, the STIS observations, and the relative size of the APO slit and of the McD slit are all illustrated in the Fig.~\ref{fig3}.  The nature of the coronagraphic spectra allows one to study the spatial origin of the Na\,\textsc{i}~D-line components.  The STIS spectrum was acquired at the orbital phase $\phi = 0.19$.

\begin{figure}
\begin{center}
\includegraphics[width=90mm]{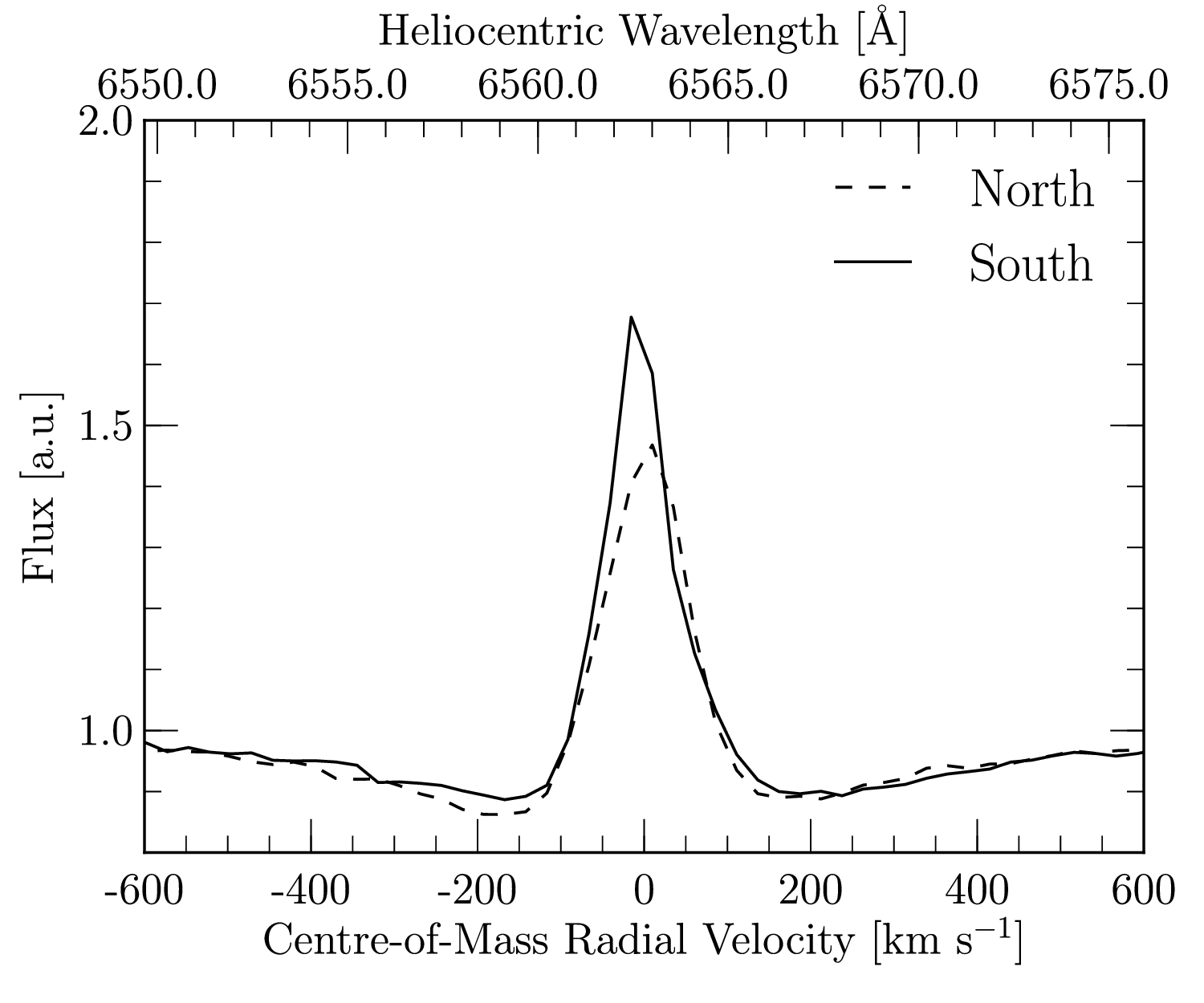}
\end{center}
\caption{Shown are the STIS H$\alpha$ line profiles for the north and south apertures extracted separately.  The resolving power of the spectra is $R = 5,000$.  To convert back to heliocentric velocity, add 19~km~s$^{-1}$.  These spectra were taken at phase $\phi = 0.17$, see Fig.~\ref{fig2}.  \label{fig8}}
\end{figure}

The 2029~s exposure of H$\alpha$, shown in Fig.~\ref{fig8}, was taken on 1998 March 26 by Theodore Gull (proposal ID~7593). The spectrum was acquired with the same aperture, grating and position angle as the Na\,\textsc{i} spectrum.  This spectrum was also obtained from the Hubble Legacy Archive.  This observation was made during the orbital phase $\phi = 0.17$.  In all STIS spectra the northern and southern regions of the slit spectrum were extracted as separate apertures with standard \textsc{iraf} routines.  The region very close to the coronagraphic bar was avoided in the aperture selection, due to contamination of the spectrum from scattering by the coronagraphic bar.  The aperture selection for north and south apertures started at about 1{\arcsec} away from the centre of the occulting bar.

\subsection{Radial velocity} \label{sec-rv}

The stellar radial velocity (RV) for each observation is required in order to remove the photospheric absorption line from the observed profile.  The removal of the photospheric absorption component from the Na\,\textsc{i}~D-line profile is discussed in {\S}~\ref{technique}.  The RV also allows us to determine the orbital parameters of the system.  The stellar RV of each observation was measured in two ways.  Method one (M1), as described below, involves measuring the centroids of three selected lines in the spectrum of the primary.  M1 is less prone to peculiarities of a single line as those would be averaged out.  However, M1 is affected by the periodically asymmetric wing present on these absorption lines.  Method two (M2) accounts for the line asymmetries by using a two-component Gaussian to deconvolve one of these lines.  Possible causes of the asymmetry are discussed in detail in {\S}~\ref{sec-lineshape}.

\begin{figure}
\includegraphics[width=90mm]{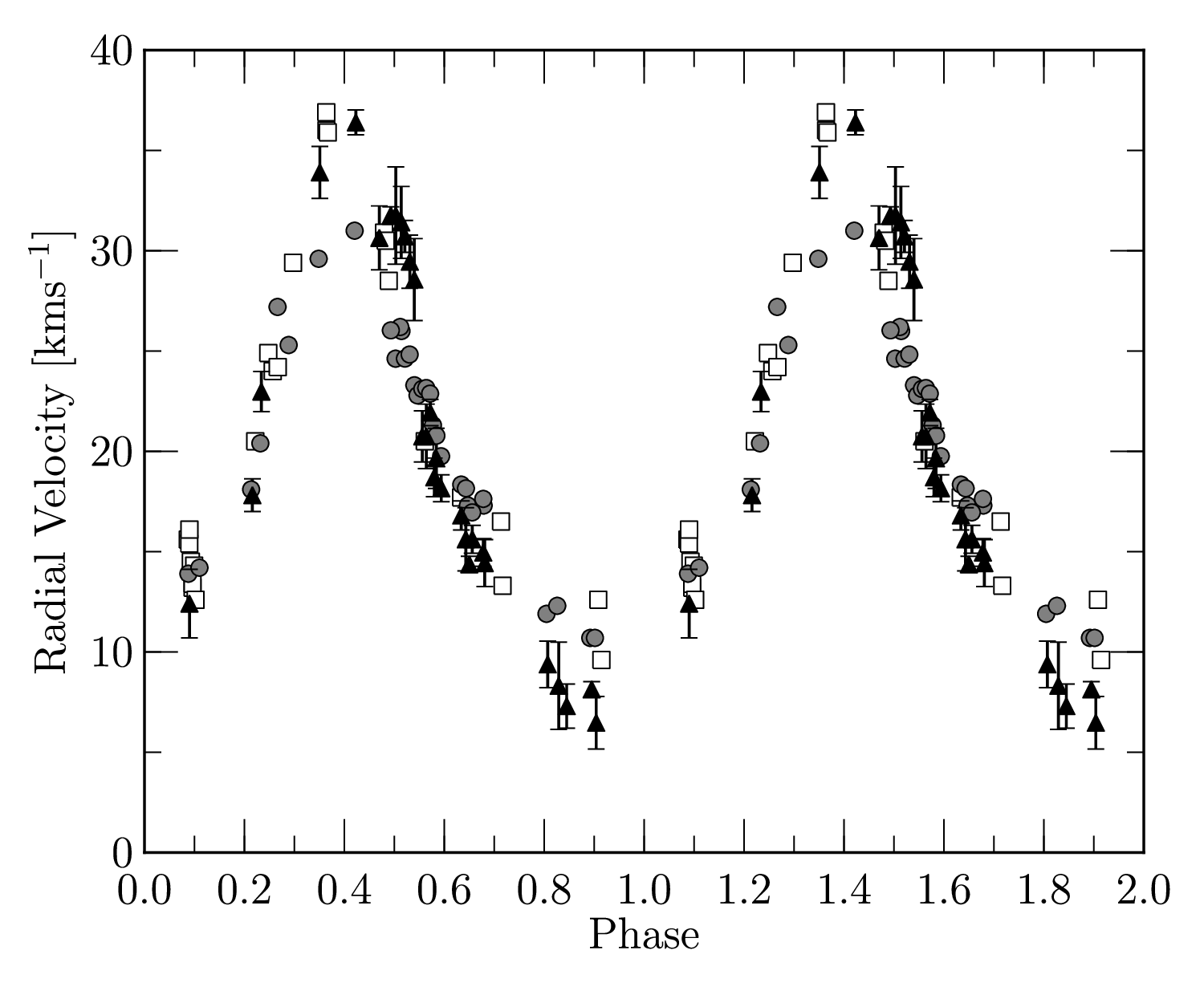}
\caption{Radial velocity curves for HD~44179, comparing the different measurements techniques.  The open squares are data from \citet{vanwinckel1995}, the grey circles are measurements made in the same manner as \citet{hobbs2004}, using M1.  The filled triangles represent the deblended strong C\,\textsc{i} component measured with M2.  See Fig.~\ref{fig2} for the location of orbital phases.  \label{fig9}}
\end{figure}

The first method (M1) was identical to the method outlined in \citet{witt2009}.  In order to construct a complete and consistent dataset for all 33 observed spectra in the APO dataset, we chose to continue using M1.  The centroid of each line, measured with \textsc{iraf}, was used to determine the RV. The three absorption lines used in those measurements were: Fe\,\textsc{ii} 4923.927~{\AA}, C\,\textsc{i} 4932.049~{\AA}, and C\,\textsc{i} 5052.167~{\AA}, and the RV results have typical random errors of $\pm$~0.4~km~s$^{-1}$ \citep{hobbs2004}.  These lines were chosen because they are not too weak to be measured, nor are they too near saturation.  These lines are also not contaminated by telluric features, they are not blends with lines from other elements, and are not unresolved blends with other multiplet lines.  The RV reported is the average of the RV determined for each of these lines.  These values are tabulated in Table~\ref{table1} under the column heading RV, and are plotted in Fig.~\ref{fig9}.  

The profile asymmetry affects the determination of the RV when using the entire profile, as in M1.  The second measurement technique (M2) uses a two-Gaussian deblending of the C\,\textsc{i} 5052.167~{\AA} line to separate the strong line core and the weaker wing.  A sample of the line profiles, shown in Fig.~\ref{fig5}, reveals that the asymmetric profiles of nearly all photospheric absorption lines vary as a function of orbital phase in the same manner.  We recognize that a two-component deblending is simplistic; however, it does allow us to quantify the asymmetric lines.  This C\,\textsc{i} line was chosen as representative of the sample of lines and met the criteria listed in the previous paragraph.  Since the lines are asymmetric during most of the orbit, this has the effect of damping the RV amplitude at maximum and minimum RV. This explains the difference between M1 RV and RV measured by \citet{vanwinckel1995} at maximum and minimum RV, as shown in Fig.~\ref{fig9}.  The FWHM of the Gaussian for both profile components was assumed to be the same, assuming the primary star is the dominant source of light in the central cavity.  The width of the symmetric profile spectrum \#\,13, which occurs at inferior conjunction ($\phi = 0.21$), was used as the intrinsic width of the underlying line profile.  From our observations, we measure a nearly constant total equivalent width having no dependence on the orbital motion of the star.  Therefore, we also constrained the equivalent width of the line to be constant.  We associate the strong component with the true RV of the primary star.

The resulting strong (RV$_{strong}$) and weak (RV$_{weak}$) components are tabulated in Table~\ref{table1}. The RV$_{strong}$ values are plotted in Fig.~\ref{fig9} (filled triangles).  The maximum RV values of M2 are in agreement with the data of \citet{vanwinckel1995}, while the minimum RV values of M1 (grey circles) match the \citet{vanwinckel1995} data better.  This indicates that M2 is at least reasonable for determining the RV of the primary star.  The results of M2 for the strong component are equivalent to measuring the RV at the deepest part of the absorption line.  The velocity of the RV$_{weak}$ with respect to the RV$_{strong}$ are plotted in panel c of Fig.~\ref{fig10} and will be discussed in \S~\ref{technique}.

\section{Orbital analysis} \label{sec-orbital}

A Lomb-Scargle periodogram \citep{scargle1982} was constructed for all 33 RV$_{strong}$ measurements presented in this paper.  The resulting period found was 317~$\pm$~2~days, which is in agreement with the 318~$\pm$~3 day period measured by \citet{waelkens1996}.  This period was used as a constraint in the \textsc{vcurve} program \citep{bertgrob1969}. 

The orbital parameters for the binary system were determined with the RV values for both M1 and M2.  M1 utilized the RV measurements listed in Table~\ref{table1} and were fit using a modified version of the \textsc{fortran} program \textsc{vcurve}.  M2 used \textsc{vcurve} to fit the strong component (RV$_{strong}$).  The orbital parameters determined from \textsc{vcurve} for both methods are shown in Table~\ref{table2}.  The velocity values from \citet{vanwinckel1995} are also presented in Table~\ref{table2}.  The quantities below the horizontal rule (a$_{1}$, M$_{T}$, and m$_{2}$) in Table~\ref{table2} are derived using the effective inclination angle of 35$\degr$ and a primary mass of 0.8~M$_{\odot}$.   Aside from the orbital period difference, the orbital parameters found via M2 are in very close agreement with those of \citet{vanwinckel1995}.  This close agreement suggests that the RV measurements using M2 produce parameters for the system are more consistent with \citet{waelkens1996}.  It should be noted that the uncertainty in the mass function and secondary mass are most sensitive to the uncertainty in the semi-amplitude.  

Based on the orbital data determined using the \textsc{vcurve} fit to the strong component (RV$_{strong}$), a to-scale schematic of the binary orbits has been constructed to aid in the interpretation of the spectra, see Fig.~\ref{fig2}.  The locations of the 33 APO observations to date are also indicated in the figure along with key orbital locations.  Orbital phase zero, $\phi = 0$, was chosen to correspond with the location of phase zero defined by \citet{waelkens1996} to aid in comparison of data.  \citet{waelkens1996} chose $\phi = 0$ to correspond to JD~2448300, near minimum RV.

\begin{figure}
\includegraphics[width=85mm]{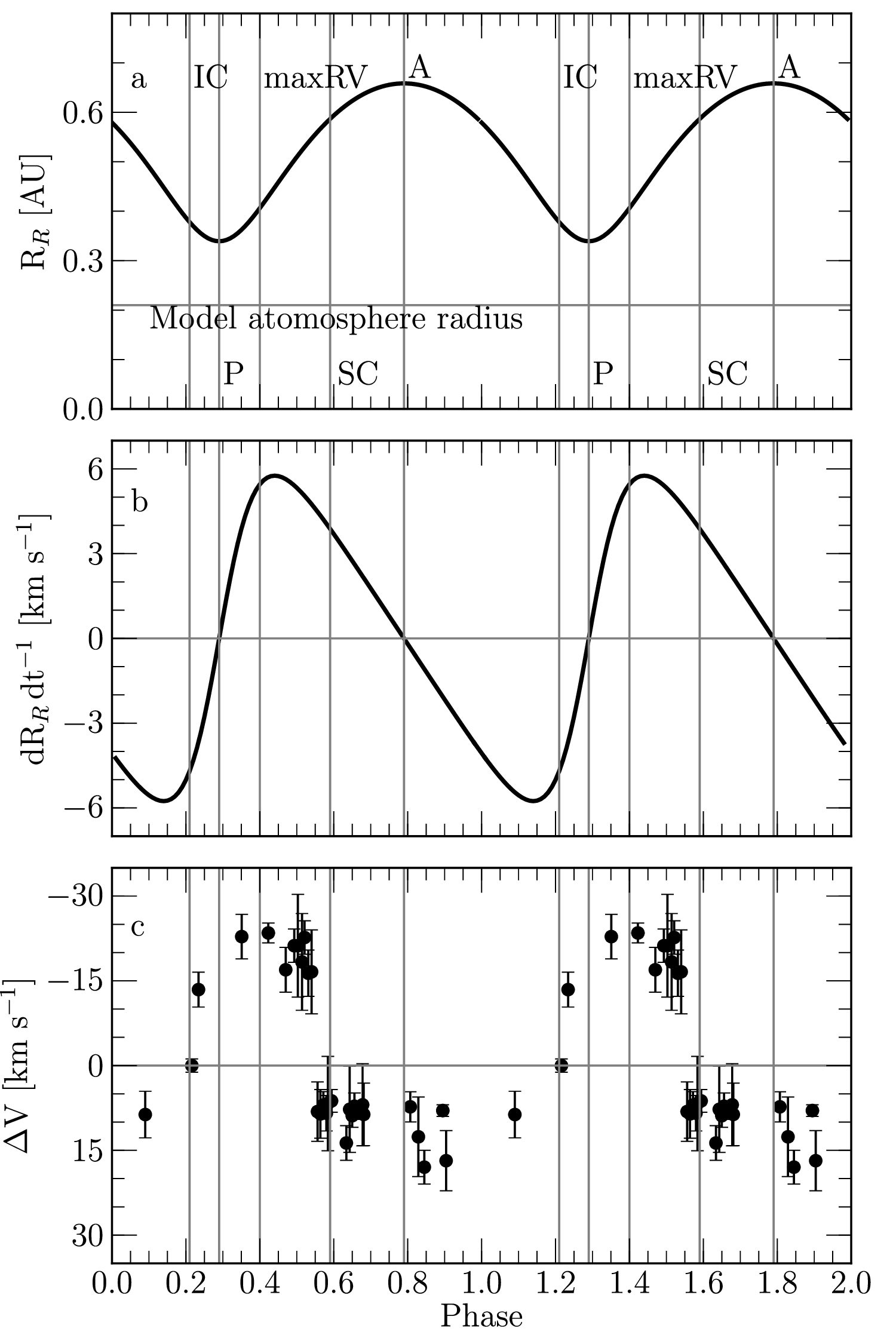}
\caption{In panel a, the Roche lobe radius in astronomical units (R$_{R}$) is plotted as a function of orbital phase as calculated by equation~(46) of \citet{sepinsky2007}.  In panel b, the derivative (dR$_{R}$~dt$^{-1}$) of the curve in panel a is shown.  Positive slope indicates an expanding Roche lobe radius, while a negative slope indicates a contracting Roche lobe radius.  In panel c, the velocity of the weak component with respect to the primary star ($\Delta$V) is plotted so that the negative velocities are above the zero-line.  Negative velocities indicate material moving away from the primary star and vice versa.   The vertical lines represent the phases of inferior conjunction (IC) $\phi = 0.21$, periastron (P) $\phi = 0.29$, maximum RV (maxRV) $\phi = 0.42$, superior conjunction (SC) $\phi = 0.59$ and apastron (A) $\phi = 0.79$. See Fig.~\ref{fig2} for the location of orbital phases. \label{fig10}}
\end{figure}

Since the components of the binary are in an eccentric orbit, the Roche lobe radius is not constant as a function of orbital phase.  The consequences of the variable size of the Roche lobe are discussed in {\S}~\ref{sec-phase_dependent-asymmetries}.  In panel a of Fig.~\ref{fig10} the Roche lobe radius of the primary star was calculated as a function of orbital phase using equation~(46) of \citet{sepinsky2007}.  The exact value of the \textit{effective} Roche radius may differ, as will be discussed later; however, the relative variation and orbital phase dependence should remain the same.

\begin{table*}
\centering
\begin{minipage}{160mm}
\caption{Orbital parameters  \label{table2}}
\begin{tabular}{llllll}
\hline
Symbol & Parameter\footnote{Orbital parameters for HD~44179 calculated with \textsc{vcurve}, uncertainties are generated by the program except for the last five parameters.  The uncertainties from \textsc{vcurve} are used to compute the uncertainties in the last five parameters. The period of 317 days was used as a starting parameter for the fit to the phase folded data.  Column M1 (method 1) and M2 (method 2) are for different radial velocity determination methods discussed in detail in \S~\ref{sec-orbital}.  }	&	M1	&	M2 & Van Winckel95\footnote{\citet{vanwinckel1995}} & Units	\\			
\hline			
P	    & period	        & 319.2 $\pm$ 0.2 	    &	318.8 $\pm$ 0.6 & 298    & days        \\
V$_{0}$	& system velocity	& 18.6 $\pm$ 0.1        &	18.4 $\pm$ 0.5 & 21.5 $\pm$ 1.2   & km~s$^{-1}$  \\
K$_{1}$	& semi-amplitude	& 9.8 $\pm$ 0.2         &	14.9 $\pm$ 0.8 & 13.0 $\pm$ 1.7  & km~s$^{-1}$  \\
e	    & eccentricity	    & 0.276 $\pm$ 0.014	    &	0.320 $\pm$ 0.041 & 0.45 $\pm$ 0.13  &      \\
$\omega$& longitude of periastron\footnote{Referenced to periastron.}	&	$-$20.1 $\pm$ 3.2	&	$+$7.3 $\pm$ 7.7 & 28 $\pm$ 16  & $^{\circ}$   \\
a$_{1}sin(i)$& semi-major axis	    &	0.281 $\pm$ 0.006	&	0.414 $\pm$ 0.081 & 0.32 & AU	\\
f(m)    & mass function     & 0.028 $\pm$ 0.002     & 0.049 $\pm$ 0.049 & 0.049 & M$_{\odot}$ \\
\hline
a$_{1}$ & true semi-major axis & 0.490 $\pm$ 0.006	&	0.722 $\pm$ 0.081 & ... & AU	\\
M$_{T}$ & total system mass & 1.49 $\pm$ 0.06      & 1.80 $\pm$ 0.37 &   ...    & M$_{\odot}$ \\
m$_{2}$ & secondary mass    & 0.69 $\pm$ 0.04      & 1.00 $\pm$ 0.37 &   ...    & M$_{\odot}$ \\

\hline		
\end{tabular}
\end{minipage}
\end{table*}

\section{PHOTOSPHERIC ABUNDANCE OF SODIUM} \label{sec-abundance}

In order to remove the photospheric absorption line from the Na\,\textsc{i}~D-line profile (see Fig.~\ref{fig4}) the Na abundance is required for the construction of a photospheric model line.  The complex line profile of the Na\,\textsc{i}~D-lines were not suitable for the determination of the Na abundance.  The near infrared (IR) photospheric absorption multiplet at 8183.256~{\AA} and 8194.824~{\AA} provided an alternate approach in the determination of the Na-abundance.  The spectral coverage of ARCES included the 8183.256~{\AA} and 8194.824~{\AA} Na\,\textsc{i} lines; however, the 8183.256~{\AA} line was buried in a forest of telluric features. Fortunately, the 8194.824~{\AA} line, the stronger of the two, fell between the telluric features for some of the observations and we were able to measure its equivalent width to be 80.7~$\pm$~3.0~m{\AA}.

The \textsc{phoenix} code, Version 15.04.00E \citep{hau1992, hau1993, hau1995, allardhau1995, baron1996, hau1996, hau1997, baronhau1998, allard2001, hau2001}, a generalized NLTE stellar atmosphere code, was utilized to compute the expected photospheric absorption line profiles for the primary of HD~44179.  The calculated profiles included Na\,\textsc{i}.  In \S~\ref{sec-lineshape} we discuss the disagreement of the calculated line profiles with the observed profiles.  The parameters adopted for this calculation were T$_{eff} =$ 7,700 K, log g $=$ 1.10, [Z/H] $= -$3.5, [CNO/H] $= -$0.5 \citep{waelkens1996}, and a stellar mass of 0.8 M$_{\sun}$ \citep{witt2009}.  The abundance determination utilized several Na\,\textsc{i} 8194.824~{\AA} line profiles with abundances from $-$0.10~dex to $-$1.00~dex.  These lines were generated with the same atmospheric parameters, with the exception of their variable sodium abundance.  Using the measured equivalent width of the Na\,\textsc{i} 8194.824~{\AA} line, a linear interpolation of the equivalent width values from the model lines was used to determine the Na abundance to be [Na/H] $= -0.18 \pm 0.05$ dex.

Sodium is not significantly depleted in the spectra of HD~44179, while Fe and other refractory elements are substantially depleted.  The peculiar abundances of HD~44179 were pointed out by \citet{vanwinckel1995}.  In our data there is no evidence of interstellar sodium in the line-of-sight to HD~44179.  The observed depletion pattern in HD~44179 is similar to that of the interstellar medium \citep{morton1974,Maas2005,Gielen2009,Giridhar2010}.  Our observation of the Na abundance is consistent with the extremely metal-poor post-AGB stars as discussed by \citet{mathis1992}.  \citet{mathis1992} and \citet{waters1992} have proposed a mechanism in which a circumbinary disc provides the most favourable conditions for `cleaning' of the atmosphere of the refractory elements.  In their model, refractory elements condense into dust and are blown away by radiation pressure.  The now refractory-depleted gas is accreted back on to the star, thus altering the atmospheric abundances.  

\section{PHOTOSPHERIC LINE SHAPES AND LIGHT PROCESSING} \label{sec-lineshape}

\subsection{Line measurements and asymmetries}

\citet{hobbs2004} noted that the unsaturated photospheric absorption lines for all species have a similar FWHM of about 38~km~s$^{-1}$.  After analysing our entire dataset we notice that not only are the FWHM for all absorption lines similar, the orbital phase dependence of the absorption line asymmetry is also the same; see Fig.~\ref{fig5} for examples. During the orbital phases of superior and inferior conjunction the photospheric line profiles are symmetric, while during the phases of maximum and minimum red-shift the line profiles are the most asymmetric.  The Na\,\textsc{i} 8194.824~{\AA} line has a FWHM of 35.6~$\pm$~4.1~km~s$^{-1}$ and the C\,\textsc{i} 5052.167~{\AA} has a FWHM of 36.9~$\pm$~1.3~km~s$^{-1}$, both in agreement with the previously measured values of \citet{hobbs2004}.  The equivalent width of the Na\,\textsc{i} 8194.824~{\AA} line is 80.7~$\pm$~3.0~m{\AA} and the equivalent width of the C\,\textsc{i} 5052.167~{\AA}  line is 167.0~$\pm$~5.7~m{\AA}.  Neither shows significant dependence on orbital phase.  While the FWHM and equivalent width show no orbital phase dependence, the line profile shapes (asymmetries) are phase dependent (Fig.~\ref{fig5}).  In the following subsection we will discuss various mechanisms, which may or may not shape the photospheric absorption lines.

\subsection{Broadening mechanisms} \label{sec-broadening}

The \textsc{phoenix} code produces lines that have the correct equivalent width, but are much narrower than the observed 38~km~s$^{-1}$.  Lines this broad are not unique among F-type post-AGB stars.  The model lines do not include any broadening mechanisms other than thermal and pressure broadening.  Significant rotation is not expected for a giant star.  While it is possible the star is rotating, rotation cannot explain the orbital phase-dependent asymmetry of the lines, as rotational broadening is symmetric. 

Macroturbulence is known to produce lines with symmetric cores and asymmetric wings that conserve equivalent width; however, the required macroturbulent velocity would need to be roughly 23~km~s$^{-1}$, which would be very supersonic for this star's atmosphere.  The isothermal sound speed is approximately 8~km~s$^{-1}$.   Macroturbulence alone cannot explain the dependence of the asymmetry on the orbital period.  It also does not explain the extent of the wing either 40~km~s$^{-1}$ blue- or red-shifted depending on the orbital phase.

The primary star is only observable via the indirect line-of-sight.  The location and nature of the scattering material is uncertain.  If the material from which the light is scattered has some intrinsic velocity dispersion, then the observed line profile will be broadened as it scatters off this material.  The scattering from the north and south of the nebula likely occur at slightly different scattering angles due to the tilt of the nebula with respect to the line-of-sight.  However, the scattering from each end of the nebula likely occurs over a range of angles of order 5$\degr$.  The fact that the binary stars are in motion around the centre-of-mass implies that the mean scattering angle with change slightly over the course of the orbit.  In order to properly explore the scattering in the RR, a 3-D Monte Carlo model is planned to further study this system. 

\subsection{Orbital phase-dependent asymmetries} \label{sec-phase_dependent-asymmetries}

A double-lined spectroscopic binary could produce the line asymmetries we observe.  The weaker component would be due to the companion star, while the stronger component would be due to the primary star. However, it is unlikely that the weaker component is due to absorption in the atmosphere of the companion star for the following reasons.  The companion star is a 0.9~M$_{\odot}$ main sequence dwarf \citep{witt2009}.  This implies that its luminosity is of order L$_{\odot}$, while the primary star has a luminosity of order 6,000~L$_{\odot}$.  The ratio of the luminosities would mean an extremely weak contribution to the line from the secondary, while the observed contribution of the weaker component to the line is between 1/3 and 1/2 of the flux of the strong component.  It is also unlikely that the weak component can be attributed to absorption in the accretion disc around the secondary.  While the accretion disc has a luminosity around 300~L$_{\odot}$, the contribution would still be smaller than 1/3 of the strong component.  In addition, the maximum effective temperature of the disc is about 23,000~K \citep{witt2009}, and one would not expect atmospheric lines from the disc to have the same ionization balance as the atmosphere of the primary, which has an effective temperature of 7,700 K.  Yet, observations of various photospheric absorption lines at different ionization states and species all exhibit the same asymmetric profiles.

Pulsations in the atmospheres of R Coronae Borealis (RCB) stars show a second absorption-line component that is shifted toward either longer or shorter wavelengths of the centre of the photospheric absorption line, producing a variable asymmetric absorption-line profile \citep{rao2008}.  These asymmetric profiles are reminiscent of the observed absorption-line profiles observed in HD~44179.  However, HD~44179 is not an RCB star.  The brightness of RCB stars can decrease by nearly 8~mag at unpredictable intervals.  HD~44179 has a very periodic and predictable behaviour and does not experience such drastic changes in brightness.  The brightness of HD~44179 changes by roughly 0.14~mag \citep{waelkens1996}.  \citet{rao2008} attribute the line doubling observed in RCB stars to shocks in their atmospheres.  Such atmospheric shocks are not limited to RCB stars, similar line splitting is also seen in RV Tauri stars.  Line splitting is also observed in other stars with loosely bound atmospheres.  Complex atmospheric motions are seen in other F-type post-AGB stars, for example SAO~96709 \citep{lebre1996}.  

We interpret the asymmetric wings as motion in the atmosphere of HD~44179.  Material in the atmosphere is systematically flowing toward the primary during some phases of the orbit and away from the primary during other phases.  When the primary moves through the phase of periastron, the asymmetry is always on the blue side of the line core.  The asymmetry is on the red side of the line core, when the primary is moving through the phase of apastron, compare Fig.~\ref{fig5} and Fig.~\ref{fig2}.  There is no discernible asymmetry at the conjunctions.  The direction of flow, relative to the primary, appears to be affected by the varying size of the Roche lobe, see panel a of Fig.~\ref{fig10}.  The Roche lobe radius of the primary at periastron is 0.34~AU.  The binary components' separation at periastron is 0.91~AU.  At apastron the stars are separated by 1.8~AU and the Roche radius of the primary is 0.66~AU.  While the model atmosphere radius of the star does not exceed the Roche lobe radius, it should be noted that radiation pressure can significantly reduce the size of the \textit{effective} Roche lobe \citep{dermine2009}.  \citet{dermine2009} showed that the radiation pressure effect becomes stronger as a giant star's effective temperature increases.  The authors only compute models for T$_{eff} \leq$ 5,000~K, but HD~44179 has a higher T$_{eff}$.  Therefore, it is likely that the effective Roche lobe radius is decreased by this effect.  The size of the Roche lobe can also be positively or negatively affected by the rotation rate of the star \citep{sepinsky2007} in the case of systems with non-synchronously rotating stars in eccentric orbits, as is the case with HD~44179.  The rotation rate of the primary star is unknown; however, significant rotation is not expected in this system.

Enhanced mass-loss occurs near periastron passage, but some of the lost material can be back-accreted onto the mass-losing star within that half the orbital period when the Roche lobe reaches its largest extent, increasing its volume about eight-fold compared to its size at periastron.  The self-accretion is likely caused by the expanding Roche lobe \citep{sepinsky2010}.  \citet{sepinsky2010} showed that for systems with initial eccentricities greater than 0.2 and a mass ratio greater than 0.006, mass overflow always leads to self-accretion.  This suggests that we may be seeing the self-accretion occurring in HD~44179.  In panel b of Fig.~\ref{fig10} we plot the derivative of the Roche lobe radius with respect to time.  Negative slopes indicate a contracting Roche lobe, while a positive slope indicates an expanding Roche lobe.  In panel c of Fig.~\ref{fig10} we show the velocity of the weak component of the asymmetric line, with respect to the star's surface.  Negative velocities indicate material moving away from the star, while positive velocities indicate material moving toward the star. 

\subsection{Light processing}

A final consideration is the uncertainty in the effective inclination angle.  While it is clear that the optically thick circumbinary disc prevents a direct view of the binary, the exact path or paths though which we observe the star affect our interpretation of the system.  The cartoon model presented in Fig.~\ref{fig1}, and in other works previously cited, contains a rather thick, flat circumbinary disc.  Such discs have been seen for example in HR~4049 \citep{dominik2003}.  If the disc is flared, we could likely see into the system with a much higher effective inclination angle.  The detailed geometry of the inner region of the RR circumbinary disc requires a 3D model that includes resonant line scattering, which is beyond the scope of the present paper.


\subsection{Technique of photospheric line removal} \label{technique}

A specific model for how the RR produces these orbital phase-dependent photospheric lines is not needed in order to remove them from the spectra.  Therefore, if we utilize a mostly empirical method to remove the absorption line component from the observed Na\,\textsc{i}~D-line profile, we will be left with the nebular contributions to the spectrum. 

Using Fourier analysis, we can construct a convolution kernel that will take model atmosphere lines produced by \textsc{phoenix} and modify them to look like the observed absorption lines in the spectra of HD~44179.  Under the assumption that the model line profiles produced by \textsc{phoenix} have the correct shape for the underlying intrinsic photosphere, the \textsc{phoenix} model Na\,\textsc{i}~D-line was taken as the `true' shape of the line.  A representative photospheric absorption line is then used as a template of the line shape for a given observation.  The observed C\,\textsc{i} 4932.049 {\AA} served as the template line shape.  The observed C\,\textsc{i} 4932.049~{\AA} line was chosen because it is not contaminated by telluric features and it is unsaturated.  We were unable to utilize the IR Na\,\textsc{i} line due to contamination of the line by telluric features during several orbital phases.

\begin{figure}
\includegraphics[width=90mm]{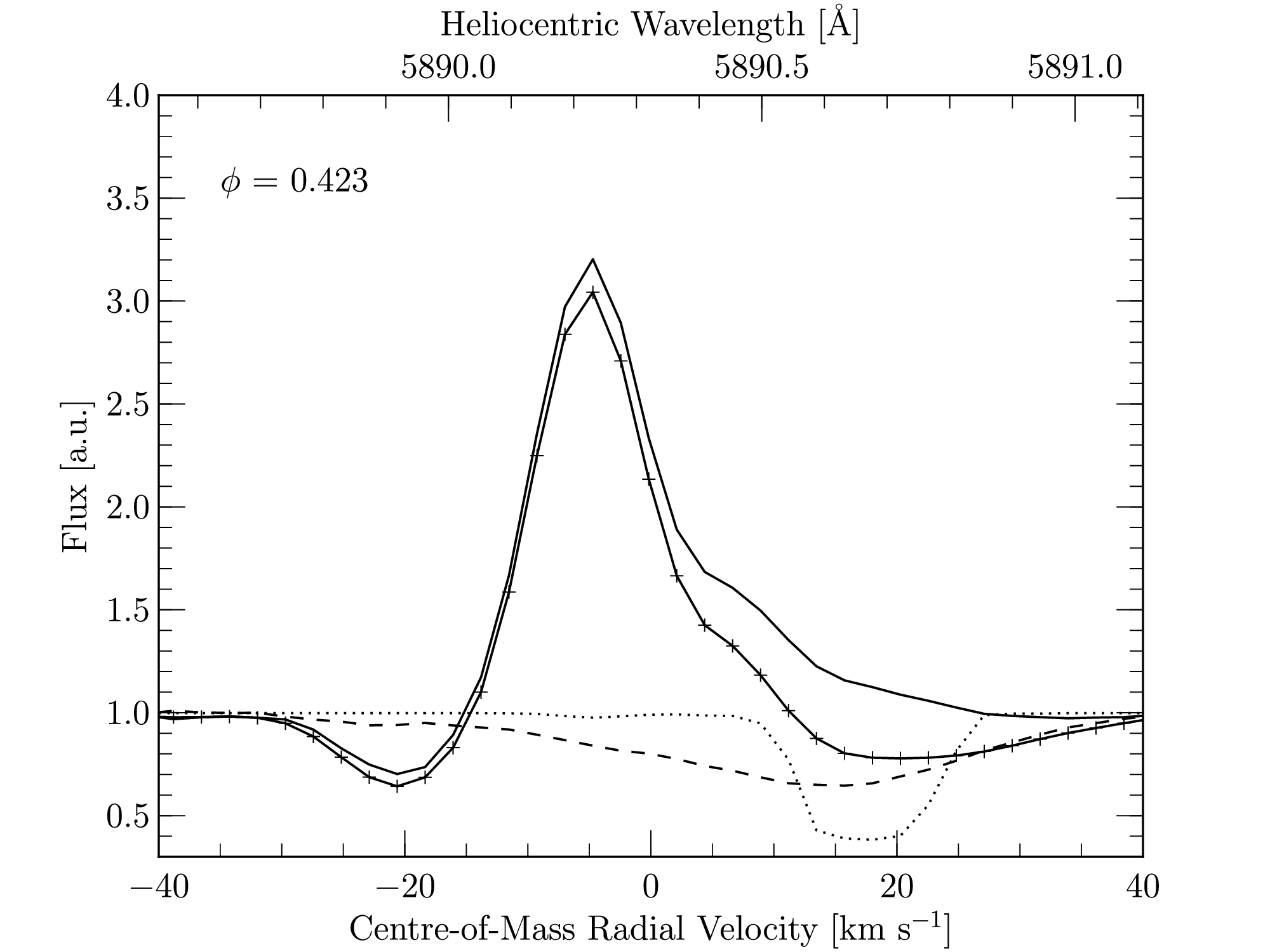}
\caption{Shown are the various steps in the model-subtraction process for the 5889.950~{\AA} line.  This representative spectrum, \#\,15, was taken at maximum red-shift of the primary, which occurs at phase $\phi = 0.423$, see Fig.~\ref{fig2}.  The normalized flux is in arbitrary units.  The dotted line is the photospheric model line calculated by the \textsc{phoenix} code.  The dashed line is the reconstructed photospheric line, which is the \textsc{phoenix} line multiplied by the response kernel constructed for the RR via Equation \ref{eq1}.  The line with $+$'s is the observed spectrum, which consists of two averaged spectra with a total exposure time of 1 hour.  The continuum S/N is approximately 300.  The solid line is the resulting model-subtracted profile.   \label{fig11}}
\end{figure}

For each observation, the \textsc{phoenix} model C\,\textsc{i} 4932.049~{\AA} line was Doppler-shifted to the appropriate wavelength according to the measured strong component of the RV.  A `response' kernel was then constructed for the RR.  The Fourier transform of the response was generated using Equation \ref{eq1}. 
\begin{equation}
\mathcal{F}\left(R\right)=\frac{\mathcal{F}\left(O_{C\,\textsc{i}}\right)}{\mathcal{F}\left(M_{C\,\textsc{i}}\right) } \ ,
\label{eq1}
\end{equation} where R is the response, $O_{C\,I}$ is the observed C\,\textsc{i} profile, and $M_{C\,I}$ is the model C\,\textsc{i} profile.  The Fourier transform of the Na\,\textsc{i}~D-line model profiles were then multiplied by $\mathcal{F}\left(R\right)$ resulting in an observationally generated reconstruction of the photospheric Na\,\textsc{i}~D-line absorption profile.  Once the reconstructed line is generated it can then be subtracted from the observed Na\,\textsc{i}~D-line profiles, thus removing the effect of the underlying stellar absorption.  An example of the \textsc{phoenix} model Na\,\textsc{i}~D-line, reconstructed line profile, and subtraction are shown in Fig.~\ref{fig11}.

\section{SODIUM D-LINE PROFILES} \label{sec-naprofiles}

The Na\,\textsc{i}~D-line profiles are a complex and variable superposition of absorption and emission components, which repeat with orbital period.  In Fig.~\ref{fig4} both observed Na\,\textsc{i}~D-lines are shown at two different orbital phases to demonstrate the extremes in the variation of the line profiles before the removal of the photospheric component.  In Fig.~\ref{fig6}, the high-resolution spectra taken at McDonald Observatory shows how distinct the various components of the line are.  The complex observed profile is a superposition of: (1) a blue-shifted (with respect to the centre-of-mass RV) absorption from the mass-loss of the primary; (2) the photospheric absorption line; (3) an emission from the northern lobe of the bi-conical nebula; (4) and an emission from the southern lobe of the bi-conical nebula.   The details of the various components and their orbital variation will be discussed in the following subsections.

\subsection{Absorption components} \label{sabsorptioncomp}

\subsubsection{Red-shifted absorption}

Perhaps the most intriguing part of the Na\,\textsc{i}~D-line profile (Fig.~\ref{fig4}) is the red-shifted absorption that first appears in observations near inferior conjunction ($\phi =  0.21$) and disappears near superior conjunction ($\phi =  0.59$).  One possible explanation for this absorption feature is infalling material along the line-of-sight to the star that is present for only part of the orbit.  If material is falling back onto the star as described in \S~\ref{sec-phase_dependent-asymmetries} it has the wrong phase dependence to explain the red-shifted absorption.  When the red-shifted absorption is present, material is actually flowing away from the star (Fig.~\ref{fig10}). A more likely explanation, therefore, is the Doppler shifted photospheric absorption line.  When the discrepancy between our \textsc{phoenix} model line FWHM and the observed FWHM is taken into account, as discussed in \S~\ref{sec-lineshape},  the stellar absorption line seems the most likely explanation for the red-shifted absorption feature.  After applying the subtraction technique in \S~\ref{technique} the effect of the underlying photospheric absorption line can be removed, as shown in Fig.~\ref{fig11}.  The process completely removes the appearance of the red-shifted absorption component of the Na\,\textsc{i}~D-line profiles for the entire range of orbital phase for which it appeared.  As a result, the infalling material explanation can be ruled out, thus leaving the Doppler shifted stellar absorption line as the explanation for the periodic red-shifted absorption feature.  However, infalling material still remains the most likely explanation of the red-winged asymmetry seen in the photospheric absorption lines during the phases interval 0.58~$\le \phi \le$~1.2.

\subsubsection{Blue-shifted absorption} \label{sec-blue-shifted-absorption}

\begin{figure}
\includegraphics[width=90mm]{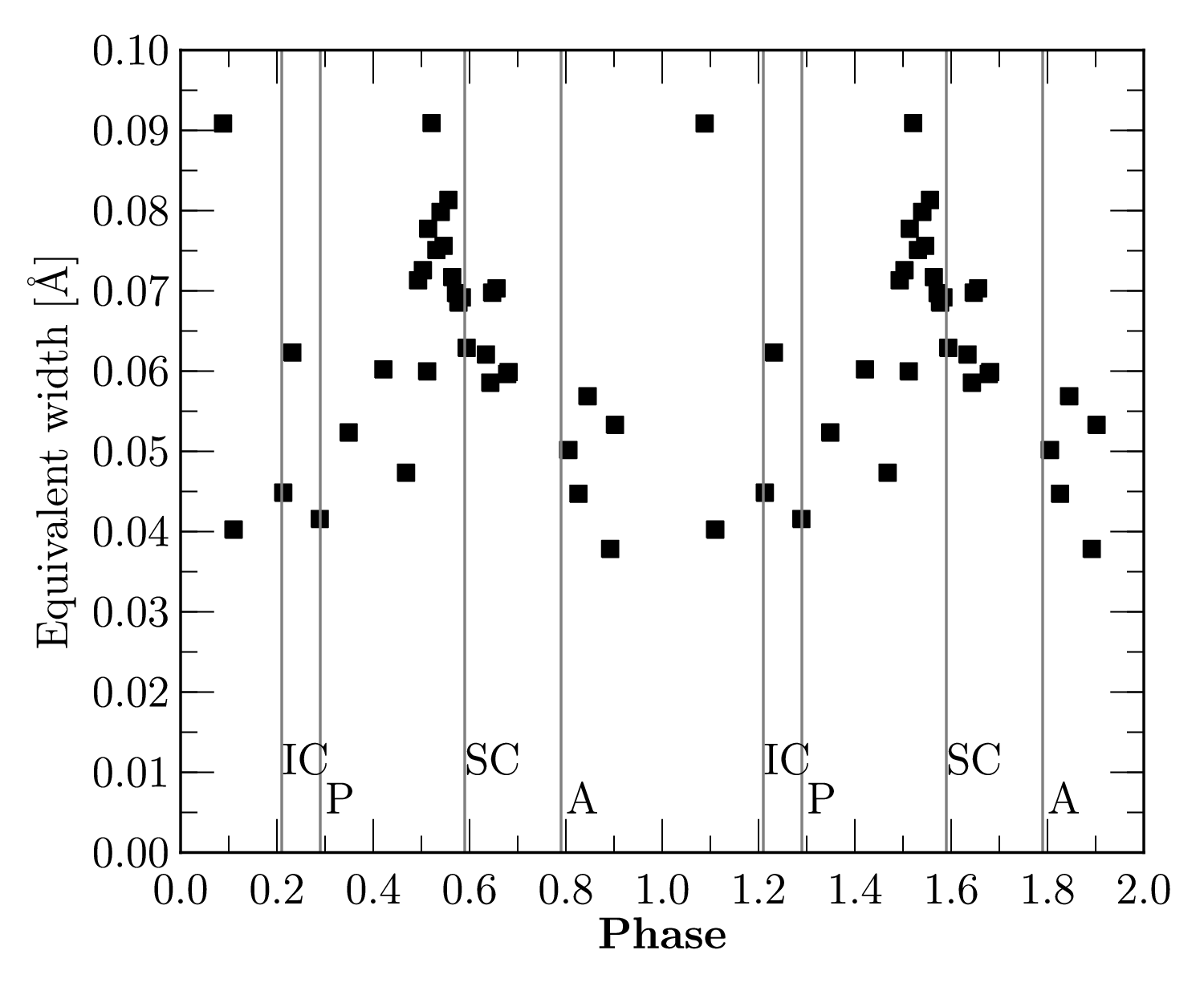}
\caption{Blue-shifted absorption equivalent width as a function of phase measured in the model-subtracted 5889.950~{\AA} line.  The uncertainty in the measurement is estimated to be about 0.01~{\AA}.  The vertical lines represent the phases of inferior conjunction (IC) $\phi = 0.21$, periastron (P) $\phi = 0.29$, superior conjunction (SC) $\phi = 0.59$ and apastron (A) $\phi = 0.79$.  Fig.~\ref{fig2} illustrates the key orbital locations.  \label{fig12}}
\end{figure}

The blue-shifted Na\,\textsc{i}~D-line absorption is present at all orbital phases, as shown in Fig.~\ref{fig4}.  The equivalent width does change as a function of orbital phase, but it is never zero, as seen in Fig.~\ref{fig12}.  The equivalent width is smallest near phase $\phi = 0.1$, with an equivalent width of about 0.04~{\AA}.  The absorption is greatest near superior conjunction ($\phi =  0.59$), with an equivalent width of approximately 0.085~{\AA}.  The time during which the equivalent width is increasing strongly coincides with the observed periodic net outflow from the primary star, as shown in panel c of Fig.~\ref{fig10}.  The material is seen in absorption against the photosphere of the primary.  Therefore, the absorbing gas must be in the indirect line-of-sight to the primary star, suggesting that the material is located in the cavity of the circumbinary-binary disc, not confined to the orbital plane.

Another constraint on the physical location of the gas is the peak velocity of the blue-shifted absorption.  Relative to the system's centre-of-mass, the velocity of the blue absorption maximum is $-$20.7~$\pm$~0.4~km~s$^{-1}$ as measured in the APO data and $-$21.4~$\pm$~0.1~km~s$^{-1}$ as measured in the McDonald data.  No dependence on the orbital phase is observed.  The terminal velocity of the blue-shifted absorption is near $-$50~km~s$^{-1}$.  This implies that the gas seen in the blue-shifted absorption against the primary star is not co-moving with the primary star, but rather moving at a fixed velocity away from the barycentre of the system, toward the observer along the line-of-sight.  Therefore, the gas seen in absorption is outside the Roche lobe of the primary and is likely gas in the general outflow.  We attribute the blue-shifted absorption to the general mass-loss driven outflow from the primary star within the central cavity of the circumbinary disk.  A lower limit of the total mass-loss rate through the bipolar outflow is given by the mass accretion rate of the disc around the secondary.  The implied accretion rate (2--5$\times 10^{-5}$~M$_{\odot}$~yr$^{-1}$), and therefore lower limit to the mass loss, is constrained by the required Lyman-continuum necessary to produce the observed free-free radio emission (\citet{jura1997} and \citet{witt2009}).

\subsection{Emission components} \label{semissioncomp}

\subsubsection{Velocities}
	
Fig.~\ref{fig4} reveals double-peaked although not fully resolved emission components in the Na\,\textsc{i}~D lines. The peaks are separated by about 12~km~s$^{-1}$ and are nearly symmetric about the system velocity.  This is confirmed by the higher-resolution spectra shown in Fig.~\ref{fig6}.  In the centre-of-mass frame the velocities remain constant with orbital phase.  This indicates that this gas is not involved in the binary motions of the star.  The blue-shifted component is Doppler shifted by about $-$6~km~s$^{-1}$ with respect to the centre-of-mass frame, while the red-shifted component is Doppler shifted by about $+$6~km~s$^{-1}$.  These double-peaked emission lines were previously attributed to material in an edge-on circumbinary disc \citep{hobbs2004}.  Material inside the cavity of the optically-thick circumbinary disc can only be viewed via the indirect line-of-sight scattering geometry proposed by \citet{waelkens1996}.  Viewing a disc at the effective inclination angle of 35$\degr$, one would expect to see a single broadened emission line from a disc, not a double-peaked profile.  We should point out that the 12~km~s$^{-1}$ full width of the CO lines observed by \citet{jura1995} is not due to the bipolar outflow; the spatial distribution of the CO emission confirms that it originates in the circumbinary disc \citep{bujarrabal2005}.

Close examination of the McDonald spectra, Fig.~\ref{fig6}, reveals that the narrow emission features sit on top of the photospheric absorption line.  The central absorption goes below the continuum, which is consistent with the location of the photospheric absorption line during that orbital phase ($\phi =$ 0.814).  The blue-shifted absorption is clearly separated from the photospheric line and the emission components.  These data also show that the emission components have shoulders.  The main emission peaks are separated by 11.6~km~s$^{-1}$.  The blue-shifted emission peak occurs at $-$4.6~$\pm$~0.2~km~s$^{-1}$ with respect to the centre-of-mass frame.  The shoulder on the blue emission occurs at $-$7.3~$\pm$~0.4~km~s$^{-1}$.  The FWHM of the blue-shifted emission is 3.2~km~s$^{-1}$.  The red-shifted emission occurs at $+$4.3~$\pm$~0.2~km~s$^{-1}$ with the shoulder occurring at $+$7.7~$\pm$~0.5~km~s$^{-1}$.  The FWHM of the red-shifted emission is 2.6~km~s$^{-1}$ and excludes the shoulder. \citet{whiteoak1983} present a single averaged spectrum of the two Na\,\textsc{i}~D-lines at $R = 120,000$; their results are consistent with our higher-resolution spectra.

Since the emission line centres are nearly symmetric about the centre-of-mass RV, this suggests the possibility that we are seeing, simultaneously, outflowing material moving both toward and away from the observer.  The Na\,\textsc{i}~D-lines are easily exited resonance lines that arise in material situated well past the inner cavity of the disc.  The coronagraphic, long-slit Na\,\textsc{i}~D STIS spectrum and the work of \citet{schmidt1991} observationally confirm that the Na\,\textsc{i}~D-line emission extends to large distances from the central source of the RR.  From the STIS spectrum, the emission can be traced to about 3.5{\arcsec} (at 710~pc this is roughly 2,500 times larger than the binary orbit) north and south of the central source.  The long-slit spectra of \citet{schmidt1991} reveal Na\,\textsc{i}~D emission as far as 10{\arcsec} north and south of the central object.  In addition, the STIS Na\,\textsc{i}~D-line spectrum shows blue-shifted emission emerging only from the southern part of the RR, while the red-shifted emission only originates from the northern part, as shown in Fig.~\ref{fig7}.  The northern view shows a weak absorption component on the blue side of the red-shifted emission close to the central source but not at larger distances from the optically-thick circumbinary disc. This is consistent with outflowing material seen against the star in the inner regions of the nebula, while at larger distances it is only seen in emission.

Current ground-based observations of the RR lack the spatial resolution of STIS, thus incorporating light from both the north and south lobes of the nebula in a single composite spectral profile.  The STIS Na\,\textsc{i}~D-line spectrum demonstrates that the double-peaked emission profile we observe from the ground originates from the extended biconical nebula.  The material in the bi-cones originates from the mass-loss driven outflow of the primary star.  The same explanation may also extend to all the other double-peaked optical emission profiles reported by \citet{hobbs2004}, but this has not been directly confirmed in the present work.  The double-peaked emission clearly does not originate from material orbiting in a circumbinary disc as previously suggested by \citet{hobbs2004}.

\subsubsection{Inclination angle}

The range in inclination angles (89$\degr$ to 79$\degr$: 90$\degr$ can be safely ruled out in light of the double-peaked emission line) leads to a range in bipolar outflow speeds from about 300 to 30~km~s$^{-1}$.  The best-fitting value of the inclination angle from \citet{bujarrabal2005}, 85$\degr$, corresponds to about 70~km~s$^{-1}$, which is of the order of the estimated escape velocity from the primary star, 83~km~s$^{-1}$.  If we assume that the outflow has a velocity equal to the escape speed from the surface of the star, then we can infer the inclination angle to be 86$\degr$.  Therefore the value for the inclination used by \citet{bujarrabal2005} would seem to fit our data best as well.  The range of velocities from 30~km~s$^{-1}$ to 70~km~s$^{-1}$, if interpreted as escape velocities, would correspond to a range in distances from the centre of the primary star of 1.6~AU to 0.3~AU respectively.  This would seem to suggest that the blue-shifted absorption, with a maximum velocity near 50~km~s$^{-1}$ ({\S}~\ref{sec-blue-shifted-absorption}), is produced by the same outflow that produces the emission lines.  Therefore, with the Na\,\textsc{i}~D-lines we observe the mass-loss driven outflow, powered by the primary star, from inside the cavity of the optically-thick circumbinary disc out as far as 10{\arcsec}.  Inside the cavity we see the material in absorption against the primary star, while outside the cavity we directly view the material in emission.

\subsubsection{Emission strength}  \label{sec-outflow}

\begin{figure}
\includegraphics[width=80mm]{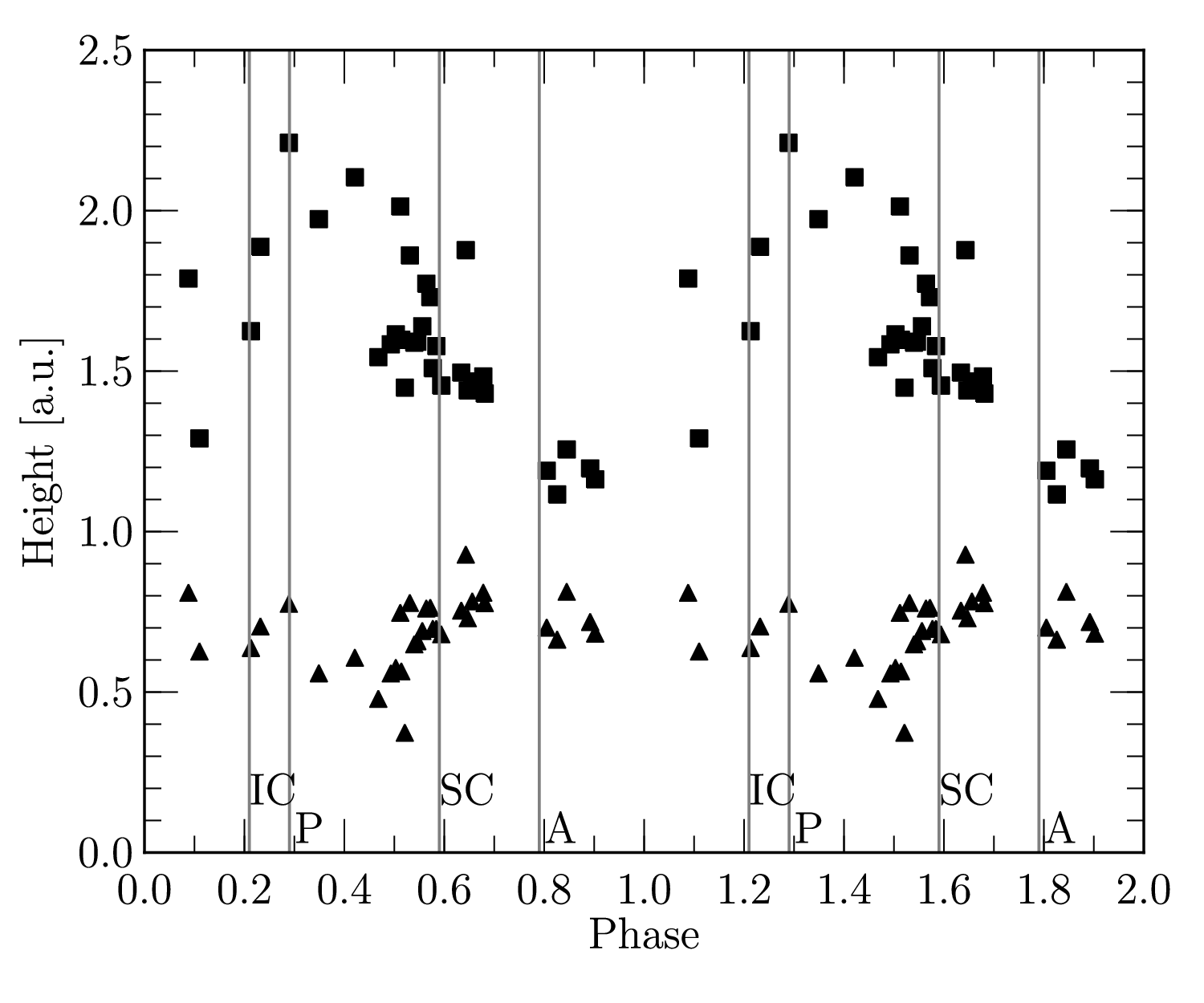}
\caption{Plotted are the V-height (squares) and the R-height (triangles) for the 5889.950~{\AA} line measured in arbitrary units as a function of phase.  The measurements have been corrected for the periodically-variable continuum level.  These measured were performed on the model-subtracted spectra.  The measurement error is on order of the size of the symbols.  The scatter in the data may represent small time scale variations and uncertainty in the continuum normalization of the data.  The vertical lines have the same meaning as in Fig.~\ref{fig12}.  Fig.~\ref{fig2} illustrates the orbital locations. \label{fig13}}
\end{figure}

The height of the centroids of the blue-shifted and red-shifted emission peaks above the continuum for both lines in the Na\,\textsc{i}~D doublet were measured as a function of phase for our model-subtracted spectra.  The height of the blue-emission peak is referred to as the V-height, while the red-shifted height is called the R-height (both are plotted in Fig.~\ref{fig13} as a function of orbital phase).  These quantities allow us to examine the components of the emission profile separately and they are preferable to using the equivalent width.  The equivalent width could in principle be used to quantify these components.  However, since the emission peaks are blended and there are absorption features on both sides of the emission peaks it complicates the interpretation of the equivalent width measurements.  The equivalent width for the blue-shifted and red-shifted emission components can only be determined through deblending, which increases the uncertainty in the measurements.  This is due, primarily, to the uncertainty of the underlying shape of each emission component.

\begin{figure}
\includegraphics[width=80mm]{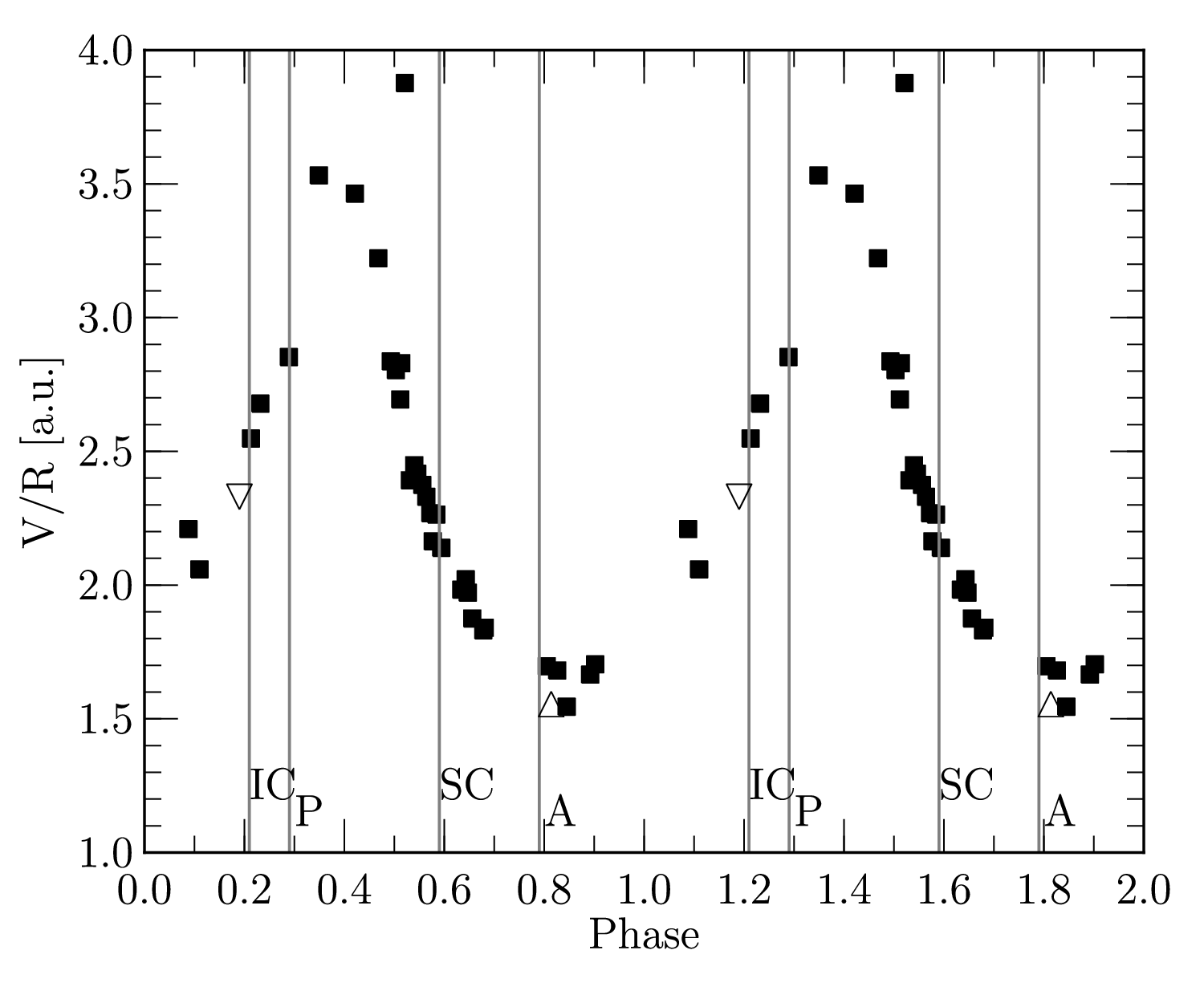}
\caption{V/R ratio for the 5889.950~{\AA} line in arbitrary units as a function of phase, measured from the model-subtracted spectra.  The scatter seen in this data as shown in Fig.~\ref{fig13} disappears because this is a relative measurement on each spectrum.  Small fluctuations on short time scales, and uncertainty in the normalization of the spectra are removed.  The measurement errors are of the order of the symbol size.  The vertical lines have the same meaning as in Fig.~\ref{fig12}. The solid squares are from the APO dataset.  The open triangle is from the artificially broadened McDonald spectrum.  The open inverted triangle is from STIS spectrum.  Fig.~\ref{fig2} illustrates the orbital locations. \label{fig14}}
\end{figure}

The outflow, as characterized by the V-height (Fig.~\ref{fig13}), is strongest near the phase of maximum RV ($\phi = 0.4$).  The R-height is greatest near phase $\phi = 0.6$.  The ratio of the V-height to the R-height (V/R-ratio) was also used to characterize the variability.   Fig.~\ref{fig14} shows the V/R-ratio for the 5890~{\AA} D2-line, which peaks near phase $\phi = 0.4$.  The V/R-ratio for both the STIS and McDonald data are included in Fig.~\ref{fig14} and show good agreement with the data from APO.  In order to make a comparison with the APO data, the McDonald datum was photosphere-subtracted and artificially degraded to match the resolution of the APO data.  The STIS data are spatially resolved and the contribution from the stellar atmosphere should be negligible due to the choice of extraction apertures.  The variability of the V/R-ratio is dominated at all phases by the blue-shifted component.  Therefore, it is not surprising that the  overall phase dependence is observed to be the same, peaking near phase $\phi = 0.4$.

The ratio of the flux between the D2-line (5890~{\AA}) and the D1-line (5896~{\AA}) for optically thin processes is the ratio of the oscillator strengths, which is equal to two.  For optically-thick processes the ratio approaches unity.  The ratio of the V-height 5890~{\AA} to V-height 5896~{\AA} is measured to be unity and shows no phase dependence.  This implies that the emission must be optically thick.  The blue absorption equivalent width (W), however, does obey the optically thin limit of W$_{5890}$~/~W$_{5896}$~=~2, which would seem to disagree with the emission ratio.  In the McDonald data the W ratio is 1.4 at its higher-resolution.  When the McDonald data were artificially degraded for comparison, the W ratio increased to 1.5.  This suggests that in the lower resolution APO data the blue-shifted emission is affecting the W ratio and that the material seen against the star is not optically thin, but not entirely optically thick.  Taking the McDonald data into consideration, it seems that we are seeing the same outflow of material in the cavity that is flowing into the outer nebula.  \citet{schmidt1991} observed that the Na\,\textsc{i}~D-line emission was strongest along the bi-cone walls or `whiskers'.  At large distances from the centre of the nebula we directly see the emission, which may be dominated by emission from walls of the outflow cavity.  The optical depth through the edge of outflow would be largest.  These observations are consistent with the work of \citet{warren-smith1981} who found that the sodium emission was optically thin on either side of the `whiskers' and optically-thick on the `whiskers'.  

Similarly, the ratio of red-shifted emission height R-height 5890~{\AA} to R-height 5896~{\AA} is also constant with phase; however, the ratio is about 0.84.  The 5896~{\AA} D1-line consistently has a stronger red component than the 5890~{\AA} D2-line.  This remains a puzzle; it was expected that the red-shifted emission component may be more highly scattered and thus weaker than the blue-shifted component.  This effect should apply equally to both lines of the doublet.  Since the ratio for the blue-shifted emission is unity, implying near optically-thick emission, the expectation would be that the red emission ratio would also be unity.  However, the smaller ratio of the red component is unexpected for the optically thick case.  This could indicate that the photospheric subtraction technique of {\S}~\ref{technique} is insufficient, but without the means to spatially separate the components at high-resolution we cannot resolve the issue in this paper.

The blue-shifted emission and blue-shifted absorption in the APO spectra are seen to be strongest near phase $\phi =$ 0.48 (just after periastron) coinciding with the period of the fastest mass outflow from the primary star, based on the asymmetry of the photospheric absorption lines, see panel c of  Fig.~\ref{fig10}.  During these phases, mass is being added to the outflow resulting in increased absorption and emission.  The close proximity in time of the strongest emission and absorption is suggestive that the outflows seen in absorption and emission are one and the same.  This argument provides further support to the discussion in {\S}~\ref{semissioncomp} relating the velocities of these components.  Note that the gas seen in Na\,\textsc{i}~D absorption is significantly slower than that of  H$\alpha$, seen in absorption near superior conjunction, and attributed to the jet from the secondary \citep{witt2009}.  This also supports the idea that the outflow seen in Na\,\textsc{i}~D emission and absorption originates from the primary star, rather than from the secondary star's accretion disc jet.  However, why the emission components, which extend to large distances from the central source, are correlated with the motion of the primary star is unknown.

\section{Single vs. double peaked emission} \label{sec-singlevsdouble}

A comparison of single-peaked emission (for example H$\alpha$ \citep{witt2009}) and double-peaked emission (as seen in Na\,\textsc{i}~D-line) reveals that they are representative of different environments in the RR system.  The STIS H$\alpha$ emission, seen in Fig.~\ref{fig8}, can be traced to roughly 1.3{\arcsec} north and south of the central source.  The H$\alpha$ has a smaller spatial extent and is seen primarily through scattering (indirect line-of-sight).  The Na\,\textsc{i}~D-line emission can be traced to 3.5{\arcsec} north and south of the central source in the STIS spectra.  The large spatial extent of the Na\,\textsc{i}~D-lines allows them to be seen both directly (at larger distances from the central source) and indirectly via scattering.  The single-peaked H$\alpha$ emission is only seen indirectly and, therefore, is arising only from the inner regions of the RR.  
The single-peaked emission lines of \citet{hobbs2004} and \citet{witt2009} most likely originate from the H\,\textsc{ii} region as mentioned by \citet{jura1997}.

In table 2 of \citet{hobbs2004} they classify 84 emission lines from 12 atomic species based on the shape of their profile.  They list 24 `narrow', 12 `broad' and 48 `double-peaked' atomic lines.  The `narrow' or single-peaked emission spectra include the [Ca\,\textsc{ii}] 7291.47~{\AA} lines.  The Ca\,\textsc{ii} 3933.66~{\AA} K-line is also among the lines classified as single-peaked, though upon close inspection an asymmetry can be seen in the line.  The spectra of \citet{whiteoak1983} reveal that, at higher-resolution, the Ca\,\textsc{ii} K-line is in fact double-peaked with a separation between peaks similar to that of the Na\,\textsc{i}~D-lines. Close examination of the double-peaked profiles reveals that the relative height of the blue-shifted and red-shifted emission peaks vary from species to species, for example the Na\,\textsc{i}~D-lines and Ca\,\textsc{ii} K-line.  Most of these, however have near-equal flux for the two components \citep{hobbs2004}.  The Ca\,\textsc{ii} K-line spectra of \citet{whiteoak1983} shows a weaker blue-shifted emission component relative to the stronger red-shifted emission component, opposite to what is seen in Na\,\textsc{i}.  The exact reason for this is unclear.  The present work did not look at the orbital phase-dependence of the other emission lines in the system.  Future work will look at how the V-height and R-height ratios of other multiplet lines behave as a function of the orbital period.  This could also be done with single lines in principle; however, with multiplet lines one assumes that each component line should be equally affected by scattering.  One also knows what the flux ratios between line components should be and also that the lines arise from the same atom; therefore, the environmental conditions giving rise to the lines are the same.  Hence, any discrepancy in such measurements may reveal more about how this complex system behaves.

\section{Conclusions} \label{sec-conclusions}

The complex profiles of the Na\,\textsc{i}~D-lines in HD~44179, periodically varying in intensity and shape with the orbital period as the photometric primary post-AGB component of the binary moves through its eccentric orbit, reveal detailed information about the varying mass-loss from the primary. Peak mass-loss occurs during the orbital phases around periastron, to be replaced by partial re-accretion during the phases near apastron.

The primary conclusions are:

\begin{enumerate}

\item The double-peaked sodium emission lines are produced in the bipolar outflow, seen via the direct line-of-sight (inclination of 85$\degr$). The observed velocity difference of 12~km~s$^{-1}$ is a result of the 5$\degr$ tilt of the bipolar outflow axis with respect to the plane of the sky.

\item The same outflow seen in the double-peaked emission is also seen in absorption against the photospheric continuum of HD~44179 via the indirect line-of-sight (effective inclination angle of 35$\degr$).  The velocity of the outflow approaching the escape velocity from the primary suggests that the primary star is the ultimate source of the outflowing material.

\item The single peaked emission lines, for example H$\alpha$, likely originate from the H\,\textsc{ii} region located in the cavity of the circumbinary disc.  The material with single peaked emission lines is at rest with respect to the centre-of-mass.

\item The red-shifted absorption component of the sodium profile, present for only part of the orbit near periastron, is most likely produced by the underlying photospheric Na\,\textsc{i}~D absorption-line of HD~44179.  The specific orbital behaviour is governed by the RV of the primary star.

\item The photospheric asymmetric blue-winged profiles in HD~44179 are consistent with enhanced mass-loss near periastron passage.  Some of the material lost is accreted onto the disc surrounding the secondary star.

\item The photospheric asymmetric red-winged profiles are consistent with self-accretion of a portion of the material lost near periastron passage.  This self-accretion occurs on a time scale of half the orbital period, centred around the phase of apastron.

\end{enumerate}

It is clear that the cause of the broadened photospheric absorption lines is of great importance for understanding and interpreting the spectra of HD~44179.  With current observations, some ambiguities cannot be resolved.  The broadening mechanisms investigated in this paper are either unlikely or cannot fully explain the observed line-profiles.  It is likely that there may be a superposition of mechanisms at play in shaping these spectral features.  The indirect line-of-sight adds to the complication of interpreting the spectra.  The effective inclination angle used here may in fact be somewhat different.  It is more likely that we see HD~44179 under a small range of angles, of which the effective inclination angle is the mean.  It is likely the case that the spread of viewing angles leads to broadened profiles.  In addition, there is a slightly different viewing geometry for the north and south ends of the nebula.  Spatially-resolved, higher-resolution spectra of the photospheric lines could help resolve some of these remaining puzzles.  A full 3-D model is called for in order to understand our complex view of the binary deep inside the circumbinary disc.

\section*{Acknowledgements}

We would like to thank the anonymous referee for the constructive comments that greatly improved this paper.  JDT would like to thank Hans van Winckel for providing a copy of \textsc{vcurve}, Dave Nero for many constructive conversations, and Johnathan Rice for assisting in the observations at McDonald Observatory.  Helpful conversations with Eric Blackman about mass exchange in close binaries are especially acknowledged.  The particular cadence of the observations presented a challenge and we thank R. McMillan and S. H. Hawley for complying with virtually all of our specific time requests each night. The APO observers provided their usual excellent assistance, allowing the most efficient completion of the program, while minimizing the impact on other observers.

\bsp \label{lastpage}


\begin{thebibliography}{99}
\bibitem[\protect\citeauthoryear{Allard 
\& Hauschildt}{1995}]{allardhau1995} Allard F., Hauschildt P.~H., 1995, ApJ, 445, 433 

\bibitem[\protect\citeauthoryear{Allard et al.}{2001}]{allard2001} 
Allard F., Hauschildt P.~H., Alexander D.~R., Tamanai A., Schweitzer A., 
2001, ApJ, 556, 357 

\bibitem[\protect\citeauthoryear{Baron et al.}{1996}]{baron1996} 
Baron E., Hauschildt P.~H., Nugent P., Branch D., 1996, MNRAS, 283, 297 

\bibitem[\protect\citeauthoryear{Baron 
\& Hauschildt}{1998}]{baronhau1998} Baron E., Hauschildt P.~H., 1998, ApJ, 495, 370 

\bibitem[\protect\citeauthoryear{Bertiau 
\& Grobben}{1969}]{bertgrob1969} Bertiau F.~C., Grobben J., 1969, RA, 8, 1 

\bibitem[\protect\citeauthoryear{Bloecker}{1995}]{bloecker1995} Bloecker T., 1995, A\&A, 299, 755 

\bibitem[\protect\citeauthoryear{Bujarrabal et 
al.}{2005}]{bujarrabal2005} Bujarrabal V., Castro-Carrizo A., Alcolea J., Neri R., 2005, A\&A, 441, 1031 

\bibitem[\protect\citeauthoryear{Cohen et al.}{2004}]{cohen2004}
Cohen M., van Winckel H., Bond H.~E., Gull T.~R., 2004, AJ, 127, 2362 

\bibitem[\protect\citeauthoryear{Dermine et al.}{2009}]{dermine2009} Dermine T., Jorissen A., Siess L., Frankowski A., 2009, A\&A, 507, 891 

\bibitem[\protect\citeauthoryear{Dominik et 
al.}{2003}]{dominik2003} Dominik C., Dullemond C.~P., Cami J., van Winckel H., 2003, A\&A, 397, 595 

\bibitem[\protect\citeauthoryear{Gielen et 
al.}{2009}]{Gielen2009} Gielen C., et al., 2009, A\&A, 508, 1391 

\bibitem[\protect\citeauthoryear{Giridhar et 
al.}{2010}]{Giridhar2010} Giridhar S., Molina R., Ferro A.~A., 
Selvakumar G., 2010, MNRAS, 406, 290 

\bibitem[\protect\citeauthoryear{Hauschildt}{1992}]{hau1992} 
Hauschildt P.~H., 1992, JQSRT, 47, 433 

\bibitem[\protect\citeauthoryear{Hauschildt}{1993}]{hau1993} 
Hauschildt P.~H., 1993, JQSRT, 50, 301 

\bibitem[\protect\citeauthoryear{Hauschildt 
\& Baron}{1995}]{hau1995} Hauschildt P.~H., Baron E., 1995, JQSRT, 54, 987 

\bibitem[\protect\citeauthoryear{Hauschildt, Baron, 
\& Allard}{1997}]{hau1997} Hauschildt P.~H., Baron E., Allard F., 1997, ApJ, 483, 390 

\bibitem[\protect\citeauthoryear{Hauschildt et 
al.}{1996}]{hau1996} Hauschildt P.~H., Baron E., Starrfield S., 
Allard F., 1996, ApJ, 462, 386 

\bibitem[\protect\citeauthoryear{Hauschildt, Lowenthal, 
\& Baron}{2001}]{hau2001} Hauschildt P.~H., Lowenthal D.~K., Baron E., 2001, ApJS, 134, 323 

\bibitem[\protect\citeauthoryear{Hobbs et al.}{2004}]{hobbs2004}
Hobbs L.~M., Thorburn J.~A., Oka T., Barentine J., Snow T.~P., York D.~G., 
2004, ApJ, 615, 947 

\bibitem[\protect\citeauthoryear{Jura, Balm, 
\& Kahane}{1995}]{jura1995} Jura M., Balm S.~P., Kahane C., 1995, ApJ, 453, 721 

\bibitem[\protect\citeauthoryear{Jura, Turner, 
\& Balm}{1997}]{jura1997} Jura M., Turner J., Balm S.~P., 1997, ApJ, 474, 741 

\bibitem[\protect\citeauthoryear{Lebre et 
al.}{1996}]{lebre1996} Lebre A., Mauron N., Gillet D., Barthes D., 1996, A\&A, 310, 923 

\bibitem[\protect\citeauthoryear{Lopez, Mekarnia, 
\& Lefevre}{1995}]{lopez1995} Lopez B., Mekarnia D., Lefevre J., 1995, A\&A, 296, 752 

\bibitem[\protect\citeauthoryear{Mathis 
\& Lamers}{1992}]{mathis1992} Mathis J.~S., Lamers H.~J.~G.~L.~M., 1992, A\&A, 259, L39 

\bibitem[\protect\citeauthoryear{Maas, van Winckel, 
\& Lloyd Evans}{2005}]{Maas2005} Maas T., van Winckel H., Lloyd Evans T., 2005, A\&A, 429, 297 

\bibitem[\protect\citeauthoryear{Men'shchikov et 
al.}{2002}]{menshchikov2002} Men'shchikov A.~B., Schertl D., Tuthill P.~G., Weigelt G., Yungelson L.~R., 2002, A\&A, 393, 867 

\bibitem[\protect\citeauthoryear{Morton}{1974}]{morton1974} Morton 
D.~C., 1974, ApJ, 193, L35 

\bibitem[\protect\citeauthoryear{Osterbart, Langer, 
\& Weigelt}{1997}]{osterbart1997} Osterbart R., Langer N., Weigelt G., 1997, A\&A, 325, 609 

\bibitem[\protect\citeauthoryear{Rao 
\& Lambert}{2008}]{rao2008} Rao N.~K., Lambert D.~L., 2008, MNRAS, 384, 477 

\bibitem[\protect\citeauthoryear{Roddier et 
al.}{1995}]{roddier1995} Roddier F., Roddier C., Graves J.~E., 
Northcott M.~J., 1995, ApJ, 443, 249 

\bibitem[\protect\citeauthoryear{Scargle}{1982}]{scargle1982}  
Scargle J.~D., 1982, ApJ, 263, 835 

\bibitem[\protect\citeauthoryear{Sepinsky, Willems, 
\& Kalogera}{2007}]{sepinsky2007} Sepinsky J.~F., Willems B., Kalogera V., 2007, ApJ, 660, 1624 

\bibitem[\protect\citeauthoryear{Sepinsky et al.}{2010}]{sepinsky2010} Sepinsky J.~F., Willems B., Kalogera V., Rasio F.~A., 2010, ApJ, 724, 546 

\bibitem[\protect\citeauthoryear{Si{\'o}dmiak et 
al.}{2008}]{siodmiak2008} Si{\'o}dmiak N., Meixner M., Ueta T., 
Sugerman B.~E.~K., Van de Steene G.~C., Szczerba R., 2008, ApJ, 677, 382 

\bibitem[\protect\citeauthoryear{Schmidt 
\& Witt}{1991}]{schmidt1991} Schmidt G.~D., Witt A.~N., 1991, ApJ, 383, 698 

\bibitem[\protect\citeauthoryear{Szczerba et 
al.}{2007}]{szczerba2007} Szczerba R., Si{\'o}dmiak N., Stasi{\'n}ska G., Borkowski J., 2007, A\&A, 469, 799 

\bibitem[\protect\citeauthoryear{Thorburn et 
al.}{2003}]{thorburn2003} Thorburn J.~A., et al., 2003, ApJ, 584, 
339 

\bibitem[\protect\citeauthoryear{Tull et al.}{1995}]{tull1995} 
Tull R.~G., MacQueen P.~J., Sneden C., Lambert D.~L., 1995, PASP, 107, 251 

\bibitem[\protect\citeauthoryear{Tuthill et 
al.}{2002}]{tuthill2002} Tuthill P.~G., Men'shchikov A.~B., Schertl D., Monnier J.~D., Danchi W.~C., Weigelt G., 2002, A\&A, 389, 889 

\bibitem[\protect\citeauthoryear{van Winckel, Waelkens, 
\& Waters}{1995}]{vanwinckel1995} van Winckel H., Waelkens C., Waters L.~B.~F.~M., 1995, A\&A, 293, L25 

\bibitem[\protect\citeauthoryear{van 
Winckel}{2003}]{vanwinckel2003} van Winckel H., 2003, ARA\&A, 41, 391 

\bibitem[\protect\citeauthoryear{van Winckel}{2007}]{vanwinckel2007} 
van Winckel H., 2007, BaltA, 16, 112 

\bibitem[\protect\citeauthoryear{Vassiliadis 
\& Wood}{1994}]{Vassiliadis1994} Vassiliadis E., Wood P.~R., 1994, ApJS, 92, 125 

\bibitem[\protect\citeauthoryear{Waelkens et 
al.}{1996}]{waelkens1996} Waelkens C., van Winckel H., Waters L.~B.~F.~M., Bakker E.~J., 1996, A\&A, 314, L17 

\bibitem[\protect\citeauthoryear{Wang et al.}{2003}]{wang2003} 
Wang X., Wang B., Pouch J., Miranda F., Fisch M., Anderson J.~E., Sergan V., Bos P.~J., 2003, SPIE, 5162, 139 

\bibitem[\protect\citeauthoryear{Warren-Smith, Scarrott, 
\& Murdin}{1981}]{warren-smith1981} Warren-Smith R.~F., Scarrott S.~M., Murdin P., 1981, Natur, 292, 317 

\bibitem[\protect\citeauthoryear{Waters, Trams, 
\& Waelkens}{1992}]{waters1992} Waters L.~B.~F.~M., Trams N.~R., Waelkens C., 1992, A\&A, 262, L37 

\bibitem[\protect\citeauthoryear{Whiteoak 
\& Gardner}{1983}]{whiteoak1983} Whiteoak J.~B., Gardner F.~F., 1983, PASAu, 5, 219 

\bibitem[\protect\citeauthoryear{Witt et al.}{2006}]{witt2006} 
Witt A.~N., Gordon K.~D., Vijh U.~P., Sell P.~H., Smith T.~L., Xie R.-H., 
2006, ApJ, 636, 303 

\bibitem[\protect\citeauthoryear{Witt et al.}{2009}]{witt2009} 
Witt A.~N., Vijh U.~P., Hobbs L.~M., Aufdenberg J.~P., Thorburn J.~A., York 
D.~G., 2009, ApJ, 693, 1946 

\end{thebibliography}
\end{document}